\documentclass{aa} 

\usepackage{graphicx}
\usepackage{txfonts}
\usepackage[citecolor=blue, linkcolor=blue, urlcolor = black, colorlinks = true]{hyperref}

\usepackage{graphicx}	
\usepackage{amsmath}	
\usepackage{amssymb}	
\usepackage{xspace}

\usepackage{siunitx}
\usepackage{comment}
\usepackage{textcomp}
\usepackage{bm}
\usepackage{lscape}
\usepackage{xcolor}
\usepackage{natbib}
\usepackage{subcaption}

\newcommand{\gdor}{$\gamma$~Dor\xspace}

\newcommand{\Msun}{\,M$_{\odot}$\xspace}

\newcommand{\xcx}{$X_{\rm c}/X_{\rm ini}$\xspace}

\newcommand{\frot}{$\Omega_{\rm nc}$\xspace}

\newcommand{\percent}{~per~cent\xspace}

\newcommand{\mesa}{\texttt{MESA}\xspace}

\usepackage{hyperref}
\begin{document}

   \title{Calibrating angular momentum transport in intermediate-mass stars from gravity-mode asteroseismology}
   \subtitle{II. Modelling 2937 BAF-type stars}
   \author{J.~S.~G. Mombarg
   \and 
   S.~Mathis
          }

   \institute{Universit\'e Paris-Saclay, Universit\'e Paris Cit\'e, CEA, CNRS, AIM, 91191 Gif-sur-Yvette, France\\
              \email{joey.mombarg[@]cea.fr}
        }

   \date{Received February 17 2026; accepted June 20 2026}
\titlerunning{Calibrating angular momentum transport in intermediate-mass stars}
\authorrunning{Mombarg \& Mathis}
 
  \abstract
   {Asteroseismology of gravity (g)-mode pulsators covering BAF-type stars have shown that angular momentum is redistributed during the main sequence. The efficiency of the transport, however, remains largely uncalibrated.   }
   {This paper aims at exploiting a sample of 2937 characterized g-mode pulsators (the largest one to-date) to place constraints on the efficiency of angular momentum transport by assuming an effective viscosity or an Eddy-viscosity based on the Tayler-Spruit dynamo within a fully-diffusive framework.  }
   {We compute grids of rotating stellar evolution models that we then use to simulate a population of stars by sampling from these grids with prior distributions on the mass, age and initial rotation rate. We then compare these simulated distributions of rotation frequencies and specific angular momentum ($J/M$) to the ones of the sample of observed stars.  }
   {We find that a fully-diffusive framework for the transport of angular momentum during the main sequence is sufficient to explain the observed evolution of near-core rotation frequencies, the observed differential rotation, and the observed mass-dependence of $J/M$ when the effective viscosity (assumed constant) is $10^6\,{\rm cm^2\,{\rm s^{-1}}}$ or larger. Viscosities predicted by the Tayler-Spruit dynamo are in general far above this value and can explain the data as well.      }
   {Future observational studies of main sequence g-mode pulsators are encouraged to measure core-to-surface rotation rates, particularly of B-type stars. In this work we have exploited the constraining potential of near-core rotation frequencies alone, while the contrast with the surface rotation would allow us to unravel the mechanisms driving the transport further.  }


   \maketitle
%
\section{Introduction}
Internal transport of angular momentum (AM) in stars can be driven by diffusive processes (magneto-hydrodynamical instabilities), advective processes (meridional circulation), a torque due to large-scale magnetic fields, and internal gravity waves (see e.g. \citealt[][]{Aerts2019-ARAA} for an overview). The magnitudes of the torques generated by each of these processes is one of the largest uncertainties in current stellar evolution models. Excited gravity (g-) mode pulsations in BAF-type dwarfs form a powerful tool to access rotation frequencies in the layer close to the boundary between the convective core and radiative envelope. \cite{Aerts2025} exploited a linear relation between the dominant pulsation frequency and the near-core rotation frequency, relying on the assumption that the dominant oscillation is a prograde dipole mode, as is the case for the majority of g-mode pulsators \citep{VanReeth2015b, Li2020, Pedersen2021}. \cite{Aerts2025-AM} presents a sample of (i) 2464 g-mode pulsators with estimated near-core rotation frequencies (model independent) by \cite{Aerts2025} and estimated masses and ages (model dependent) by \cite{Mombarg2024-gaia}, and (ii) 490 F-type g-mode pulsators with estimated near-core rotation frequencies by \cite{Li2020} and estimated masses and ages by \cite{Fritzewski2024-gdor}. The sample covers stars of masses ranging from $\sim$1.3\Msun to 8.8\Msun.

\cite{Aerts2025-AM} found a `break' in the specific angular momentum $J/M = \frac{2}{3} R^2 \Omega$ when plotted as a function of stellar mass, where the rotation frequency $\Omega$ is assumed to be constant throughout the star. Stars with masses $\lesssim 2.5$\Msun show a steeper mass-dependence for $J/M$ compared to stars above this mass. She suggests the break could be the result of weak magnetized stellar winds that apply an external torque on the stars below the break. However, prescriptions for mass loss rates of low-mass stars have only been calibrated up to $\sim$1.3\Msun \citep{Matt2015}. Additionally, stronger internal dynamo-driven magnetic fields in A- and F-type stars \citep{Brun2005} compared to B-type stars can drive stronger internal AM transport. Furthermore, \cite{Aerts2025-AM} observes an increase in the specific AM with age for stars with masses $\gtrsim 2.5$\Msun, which suggests that these stars develop differential rotation. 

For a sample of 52 slowly pulsating B-type (SPB) stars, \cite{Pedersen2022b} has provided evidence that AM is being rearranged throughout the stars, investigating cases with no AM transport, instantaneous AM transport, and AM transport driven by a Tayler-Spruit (TS) dynamo, all for an initial rotation rate of 10\percent of the critical rotation frequency. Similarly, \cite{Moyano2023} find that models including the TS dynamo better match asteroseismically derived near-core rotation frequencies of F-type pulsators ($\gamma$\,Doradus) compared to models that include only hydrodynamical instabilities. Slowly rotating \gdor stars that are situated close towards the end of the main sequence with measured near-core and surface rotation frequencies have been used by \cite{Mombarg2023-am}, the first paper of this series, to calibrate the efficiency of AM transport in a fully-diffusive framework. This calibration was done by assuming a constant effective viscosity or calculating the viscosity from (magneto)hydrodynamical processes \citep{Heger2000, Heger2005}. 

In this paper, we expand this effort from F-type stars up to B-type stars using the sample of \cite{Aerts2025-AM}, currently the largest sample of characterized g-mode pulsators, to investigate whether stellar models with only diffusive AM transport can explain the observed distributions and what constraints can be placed on the efficiency (viscosity) of the transport. This sample contains stars that cover a large range of critical rotation rates according to the measured distributions (see Fig. 2 of \cite{Aerts2025-AM}).

\section{Method}
In this paper, we use the sample of \cite{Aerts2025-AM} containing 2937 BAF-type g-mode pulsators with measured near-core rotation frequencies. The number of stars is slightly smaller than the sum of the two samples mentioned in the Introduction due to duplicate stars and outliers. 
Previous studies that aimed at placing constraints on the efficiency of AM transport typically do this by choosing the most extreme initial conditions such that the rotation velocities of the sample (or some percentile) are contained within the fast and slow rotating models \citep[e.g.][]{Gallet2013, Ouazzani2019, Li2020, Pedersen2022b, Moyano2023}. Here, we aim to model the entire observed distributions of the rotation frequencies. This additionally requires a calibration of the initial conditions. 

\begin{figure}[h]
    \centering
    \begin{subfigure}{0.49\textwidth}
        \centering
        \includegraphics[width=\linewidth]{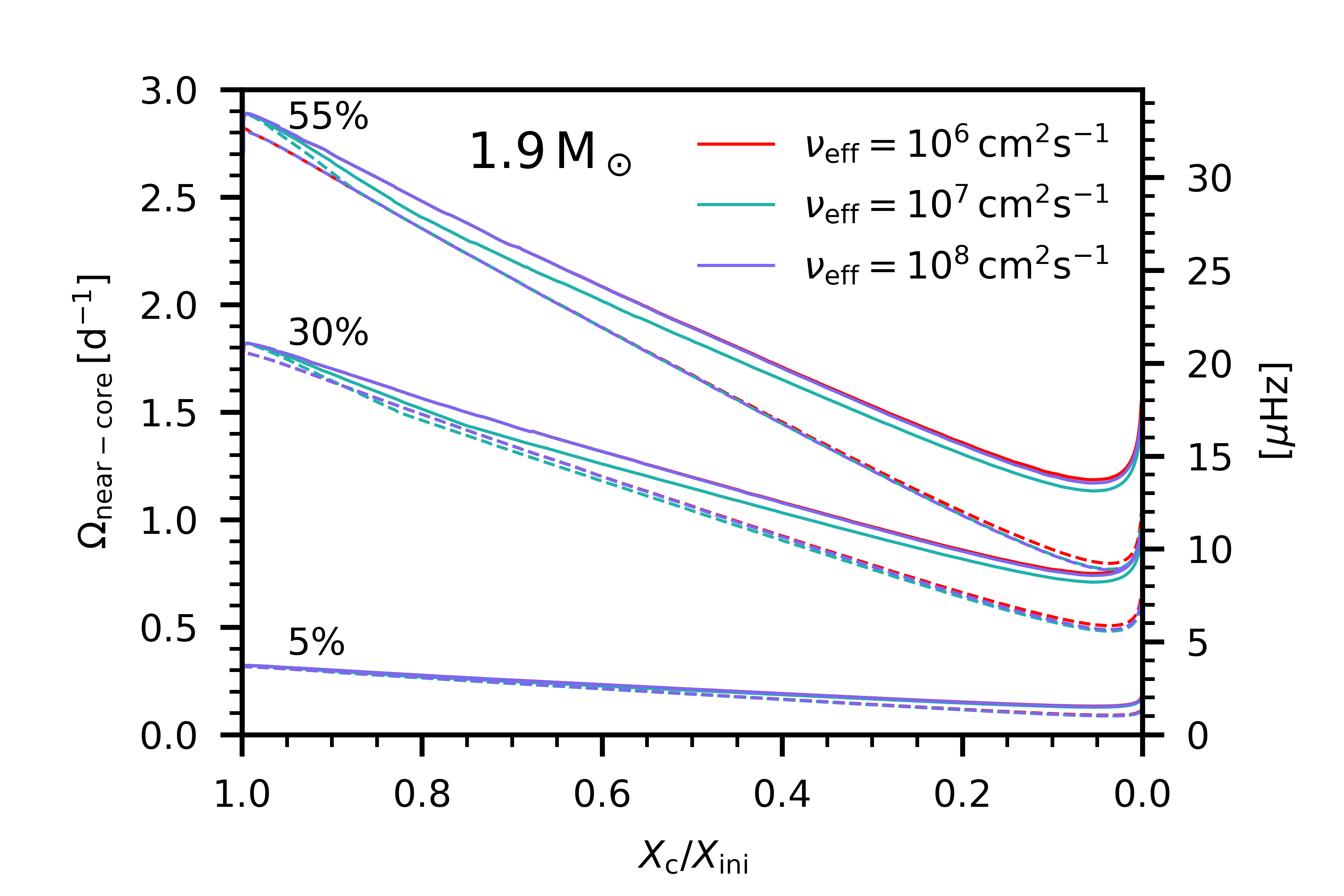}
        \label{fig:sub1}
    \end{subfigure}
    \hfill
    \begin{subfigure}{0.49\textwidth}
        \centering
        \includegraphics[width=\linewidth]{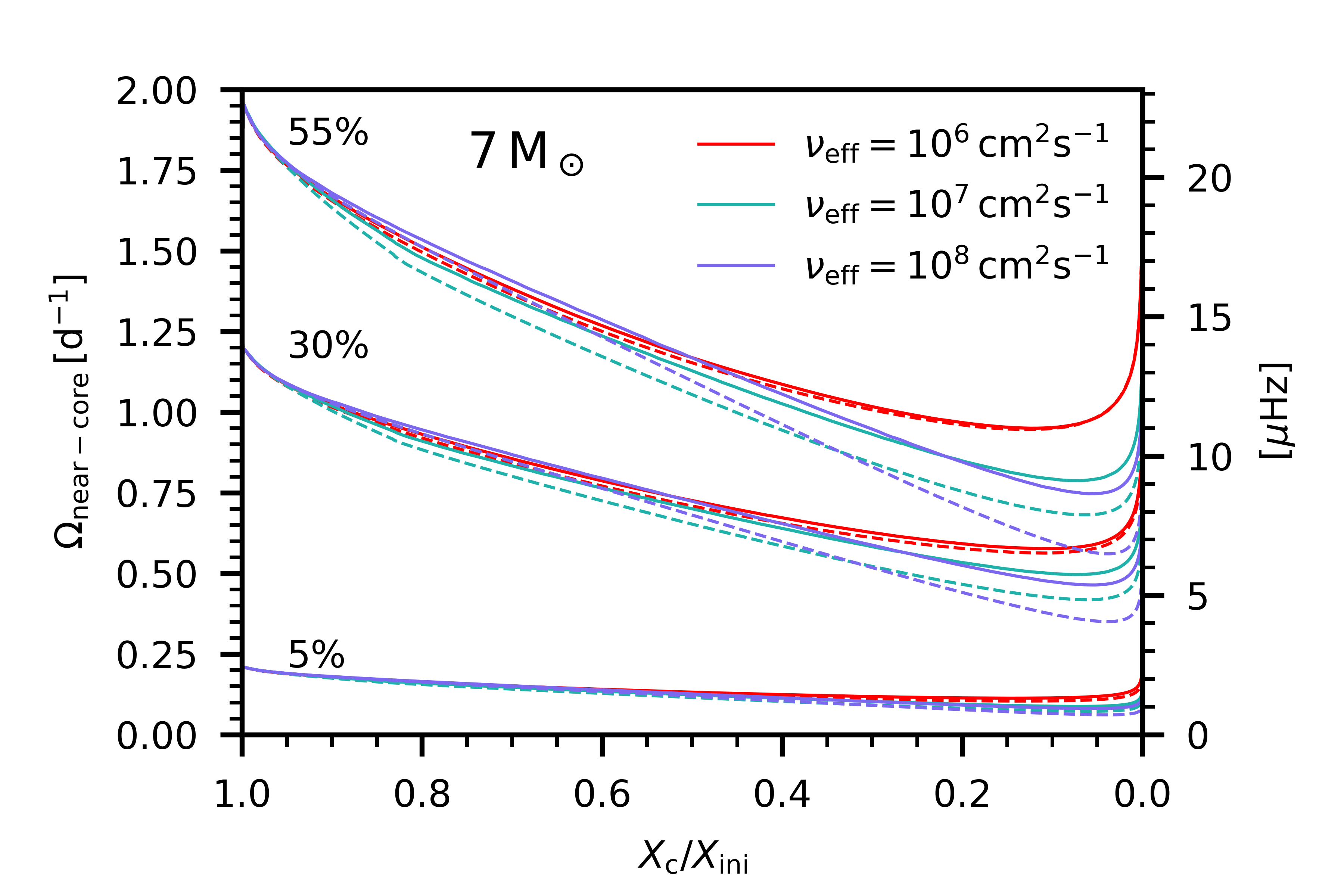}
        \label{fig:sub2}
    \end{subfigure}
    
    \caption{Evolution of the near-core rotation frequency for initial rotation rates of 5, 30, and 55\percent of the critical rotation frequency. Solid lines correspond to models with $f_{\rm CBM} = 0.005$, dashed lines correspond to models with $f_{\rm CBM} = 0.025$.  }
    \label{fig:omega_evol}
\end{figure}

\subsection{Initial conditions}
To simulate a population of g-mode pulsators, we require priors on the initial mass, the initial rotation rate, and the efficiency of the core-boundary mixing (CBM, or convective core overshoot). We rely on the grid of stellar structure and evolution models computed by \cite{Mombarg2024-gaia} with \mesa \texttt{r24.03.1} \citep{Paxton2011, Paxton2013, Paxton2015, Paxton2018, Paxton2019, Jermyn2023}, and compute additional grids with the same physics. The rotation is initialized at the ZAMS, starting from solid body rotation, where the angular velocity, $\Omega$, is set to a specified fraction of the Keplerian critical rotation frequency. We do not account for any AM loss due to stellar winds. For the CBM, an exponentially decaying diffusion constant is assumed \citep{Freytag1996},
\begin{equation}
    D_{\rm CBM}(r > r_0) = D_0\exp\left({\frac{-2(r-r_0)}{f_{\rm CBM}H_{P,0}}}\right),
\end{equation}
where the scale height is based on the local pressure scale height ($H_P$), multiplied by a factor $f_{\rm CBM}$. The subscript `0' indicates quantities evaluated at the convective interface. The grids cover masses from 1.3 to 9\Msun (step size between 0.1-1\Msun), CBM $f_{\rm CBM} \in [0.005, 0.015, 0.025]$, and initial rotation rates from 0.05 to 0.55 times the initial Keplerian critical rotation frequency, 
\begin{equation}
   \omega_0 \equiv (\Omega_{\rm surf}/\Omega_{\rm crit, Kep})_{\rm ZAMS}.  
\end{equation}

The work of \cite{Mombarg2024-distr} provides distributions for the mass, initial rotation, and CBM efficiency based on 539 \gdor stars observed with the \textit{Kepler} mission. The work shows that the mass distribution of \gdor stars follows a more or less Gaussian distribution. This is because g-mode pulsations in \gdor stars are only expected to be excited during a part of the main sequence \citep[e.g.][]{dupret2004}. For the more massive ones this is during the later part, while for the lower mass ones it is the other way around. Furthermore, this Gaussian-like distribution has also been found in the \gdor mass regime using $\sim 14,000$ g-mode pulsators observed with Gaia and TESS \citep{Mombarg2024-gaia}. The higher-mass SPB pulsators, however, seem to follow a Salpeter initial mass function (${\rm d}N/{\rm d}M \propto M^{-2.35}$).

The distribution of the initial rotation rate (as a fraction of the critical one) is also found to be similar to the Gaussian distribution centred around $\sim 0.25$ by \cite{Mombarg2024-distr}. Lastly, the probability distribution of $f_{\rm CBM}$ decreases linearly with the value of $f_{\rm CBM}$, where a value of 0.005 is roughly twice as likely as a value of 0.035. In this work, we assume that these distributions also hold for SPBs, and use them as priors to construct a synthetic population. 

In the fully-diffusive framework assumed here, the local AM transport is described by,
\begin{equation}
    \rho \frac{{\rm d}(r^2 \Omega)}{{\rm d}t} = \frac{1}{r^2} \frac{\partial }{\partial r}\left( \rho \nu_{\rm eff} r^4 \frac{\partial \Omega}{\partial r} \right), 
\end{equation}
where $\nu_{\rm eff}$ is the effective viscosity. Here, ${\rm d}/{\rm d}t$ is the Lagragian derivative, $\partial_t + \dot{r}\partial_r$. To facilitate future comparisons with schemes that are not fully diffusive, we can relate the effective viscosity to an AM flux, $F_J$, via,
\begin{equation}
     F_J =  - \frac{2}{3}  \rho \nu_{\rm eff} r^2 \frac{\partial \Omega}{\partial r}.  
\end{equation}
We assume shellular rotation, where the local rotation frequency depends only on the stellar radius \citep{Zahn1992}.
\subsection{Angular momentum transport}
In this paper we focus on g-mode pulsators with measured near-core rotation frequencies. As shown in Fig. 10 of \cite{Mombarg2023-am} for a 2\Msun star, the near-core rotation frequency will start to corotate with the convective core for effective viscosities above roughly $10^5\,{\rm cm^2\,s^{-1}}$. Therefore, the evolution of the near-core rotation does not change when increasing the viscosity beyond this value. Moreover, to reproduce the observed near-uniform rotation profiles of \gdor stars found by \cite{VanReeth2018}, \cite{Li2020}, and \cite{Saio2021}, effective viscosities higher than $2 \cdot 10^5\,$-$\,5 \cdot 10^7\,{\rm cm^2\,{\rm s^{-1}}}$ are needed. A similar conclusion was reached previously by \cite{Ouazzani2019} based on the near-core rotation frequencies alone. The AM transport in the grid of \cite{Mombarg2024-gaia} that we are using here is also done in a fully diffusive description, assuming a constant effective viscosity of $10^6$, $10^7$ or $10^8\,{\rm cm^2\,{\rm s^{-1}}}$. 

The observational limits are illustrated in Fig.~\ref{fig:omega_evol}, where we show the effect of the viscosity on the evolution of the near-core rotation frequency, \frot. As can be seen in the top panel, for F-type stars any viscosity higher than order $10^6\,{\rm cm^2\,{\rm s^{-1}}}$ will not change the evolution. For the higher-mass B-type stars (bottom panel), the limit is on the order of $10^7\,{\rm cm^2\,{\rm s^{-1}}}$. Additionally, we also investigate AM transport via the TS dynamo as implemented in \mesa \texttt{r24.03.1} (and earlier versions) following \cite{Spruit2002} and \cite{Heger2005}. The viscosities predicted from the TS dynamo are typically several orders of magnitude larger than the effective viscosities covered in this work for the stars studied here and should thus be able to drive the required AM transport.

\subsection{Simulating a stellar population}
As mentioned, we rely on previous work of \cite{Mombarg2024-distr} who derived a distribution for the CBM efficiency of a large sample of g-mode pulsators. We use this distribution as our prior on the CBM efficiency,
\begin{equation}
      f_{\rm CBM} \sim p(f_{\rm CBM}) = 43 - \frac{20}{0.035}f_{\rm CBM}. 
\end{equation}
We assume a normal distribution of initial rotation rates (as a fraction of the initial critical rotation frequency) following \cite{Mombarg2024-distr} and we leave the mean $\overline{\omega_0}$ and standard deviation $\sigma_{\omega_0}$ as free parameters to be calibrated to the observed sample,
\begin{equation}
    \omega_0 \sim \mathcal{N}(\overline{\omega_0}, \sigma_{\omega_0}).
\end{equation}
We sample the masses and ages according to the observed distributions. For the ages, we use the normalized hydrogen-mass fraction in the core, \xcx, as a proxy. We generate a population of 1000 stars with the stellar parameters drawn from these probability density functions (PDFs). While interpolation between stellar evolution tracks with conditional normalizing flows has been proven to be possible, interpolation of the stellar age remains challenging with the setup of \cite{Mombarg2024-gaia}. Therefore, we simply select the closest stellar model for each of the 1000 simulated stars and we do a linear interpolation of the relevant quantities between two \mesa time steps. For a given target age, using \xcx, we compute the rotation frequency and radius at that point for each of the 1000 simulated stars and perform a Gaussian kernel density estimate (KDE), using the Silverman method \citep{Silverman1986}, on the resulting distributions for \frot and $J/M$.     

\begin{figure}
    \centering
    \centering
    \begin{subfigure}{0.49\textwidth}
        \centering
        \includegraphics[width=\linewidth]{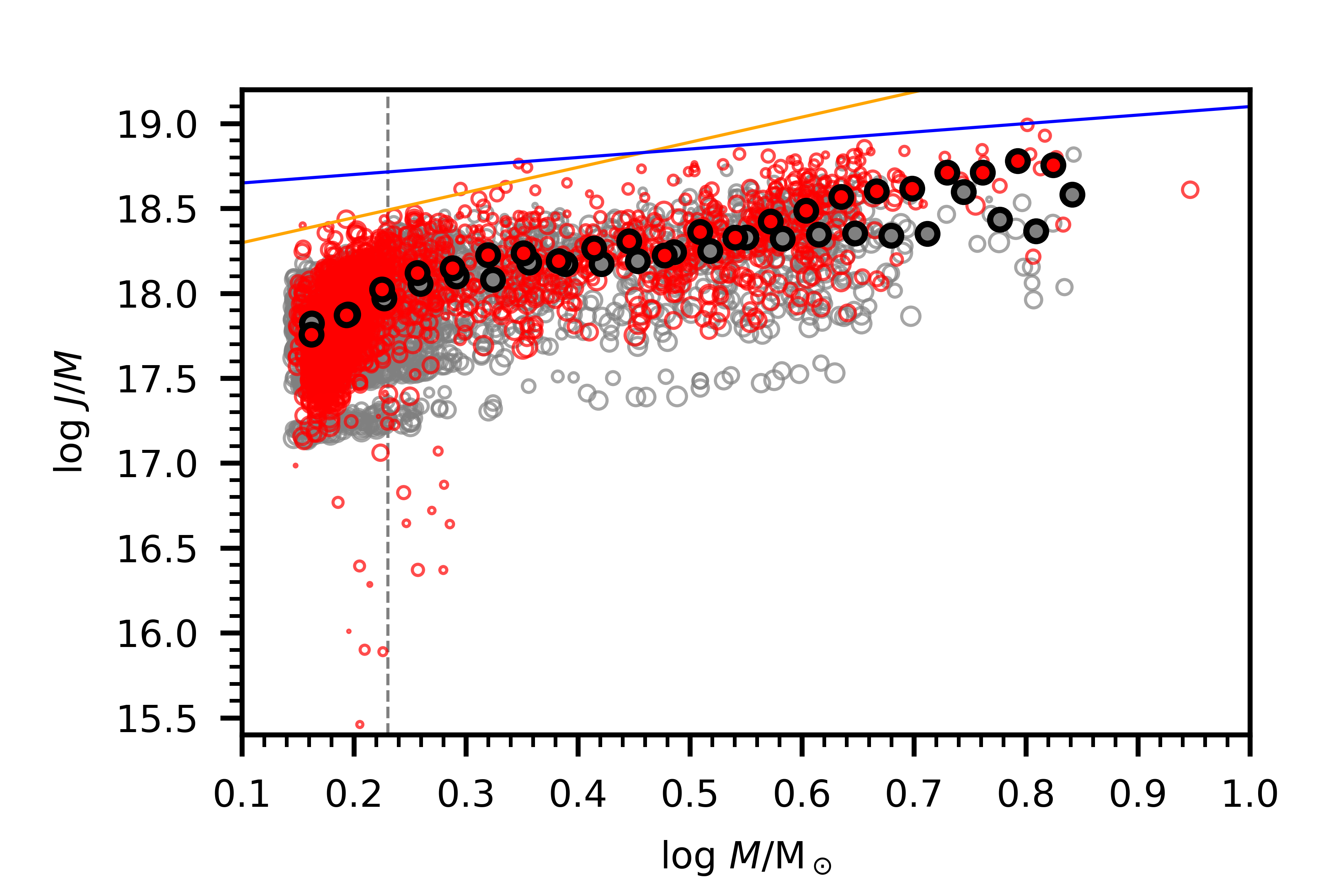}
        \label{fig:sub1}
    \end{subfigure}
    \hfill
    \begin{subfigure}{0.49\textwidth}
        \centering
        \includegraphics[width=\linewidth]{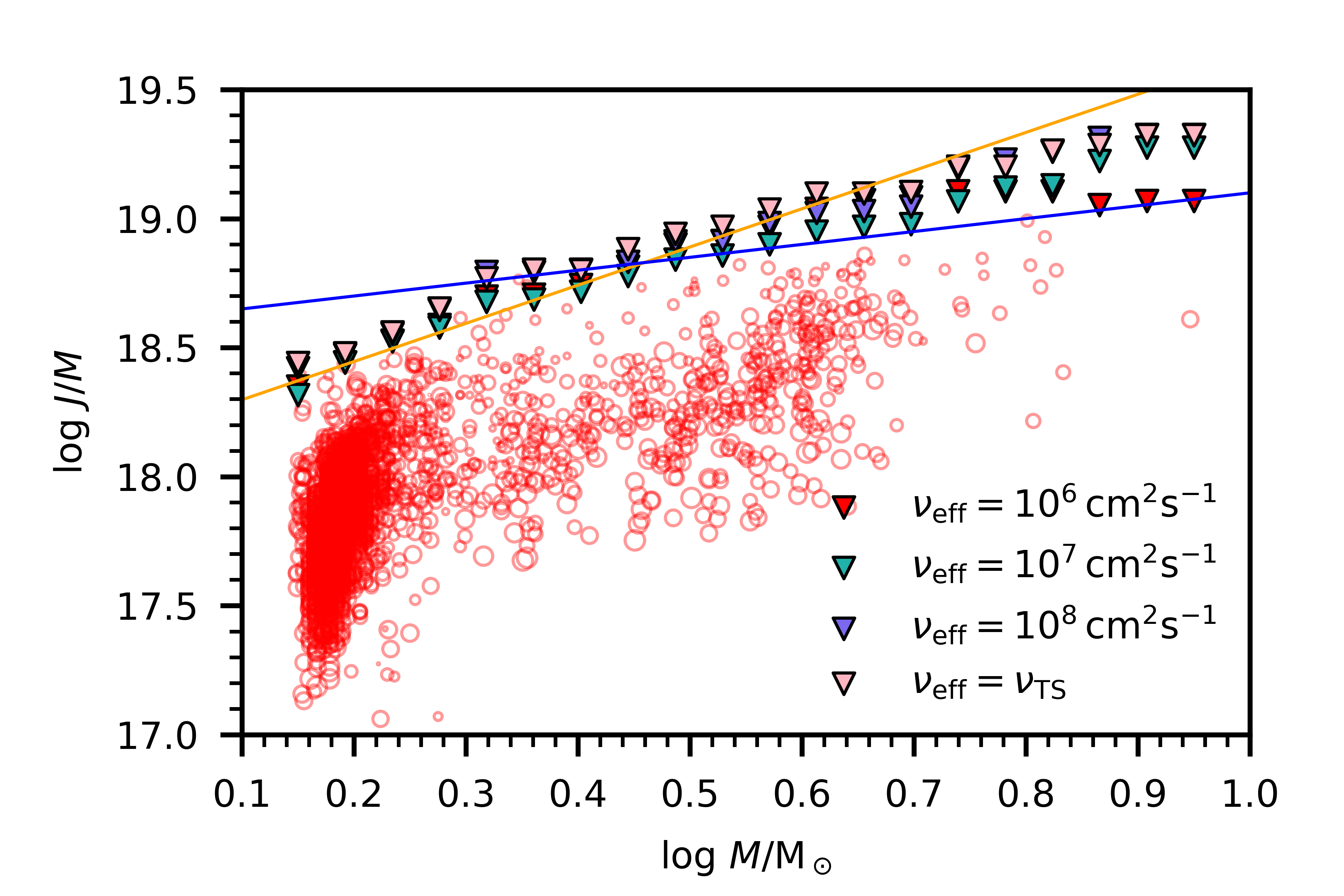}
        \label{fig:sub2}
    \end{subfigure}
    \caption{Specific AM as a function of stellar mass. The red symbols correspond to the sample of \cite{Aerts2025-AM}, the grey symbols correspond to the simulated sample of 3000 stars. The symbols with black outlines indicate the bin average. The blue and orange lines indicate the upper limits given by \cite{Aerts2025-AM}. The vertical grey dashed line indicates 1.7\Msun, diving the two mass ranges with different priors on the initial rotation rate. The symbol size scales linearly with \xcx. The bottom panel shows the upper limits in the grids for different values of the viscosity.}
    \label{fig:logJM}
\end{figure}

\section{Modelling mass dependence of $J/M$} \label{sec:modelling}
As mentioned, the study by \cite{Aerts2025-AM} shows that the mass dependence of the specific AM, estimated as $\frac{2}{3} R^2 \Omega_{\rm nc}$, changes around 2.5\Msun, where the decrease in $J/M$ with decreasing mass becomes steeper below this mass. This raises the question whether the lower-mass stars that still have a small convective outer layer experience some external torque. Here, we investigate whether the observed $J/M$ values can be explained without the need for any external torques.

We simulate a population of 3000 stars using the observed mass and \xcx distributions as priors. As mentioned before, we assume that the initial rotation rates (expressed as a fraction of the Keplerian critical rotation frequency at the ZAMS) follow a normal distribution \citep{Mombarg2024-distr}, where we calibrate the mean and standard deviation of the distribution to the observations. As is shown in \citet[][their Fig. 12]{Li2024} the measured rotation frequencies of g-mode pulsators in the young open cluster NGC2516 are nearly constant as a function of effective temperature down to $\sim7600$\,K. Below this temperature, a drop in the near-core rotation frequency is observed (for a handful of stars in total). Here, we find that stars below $\sim$1.7\Msun require a different initial distribution of the rotation rate $\omega_0$, with a lower average than the stars more massive than 1.7\Msun. We take $(\overline{\omega_0},\sigma_{\omega_0}) = (0.2, 0.1)$ for $M \le 1.7$\Msun and $(\overline{\omega_0},\sigma_{\omega_0}) = (0.3, 0.25)$ for $M > 1.7$\Msun. Although the exact choice to have two mass regimes with a division at 1.7\Msun is somewhat arbitrary, we find indications that stars below roughly 1.7\Msun undergo a different rotational evolution during the pre-main sequence (e.g. AM loss via magnetized winds) that slows them down compared to stars more massive than this.  

Figure~\ref{fig:logJM} shows the specific AM of the observational sample and the simulated sample. In general, the simulated data are in agreement with the observational data with some deviation at the higher mass end, where the observations are sparse. We note that no stars in the simulated population are situated above the upper limits found by \cite{Aerts2025-AM}, indicated by the yellow and blue lines in Fig.~\ref{fig:logJM}. Furthermore, the simulated population contains a few stars above $\log M = 0.3$ with lower $J/M$ than the lower limit of the sample. These are models with an initial rotation rate of $\omega_0 = 0.05$. The ridge seen in the simulated data (grey points) is due to the discrete step size in the initial rotation rate. When we compare distributions, the use of KDEs will smooth out these discrete steps. 

Looking at the physical upper limits on $J/M$ of the models in the grid, we find that stars exist above the upper limit given by \cite{Aerts2025-AM} for $M > 2.5$\Msun, as shown by the triangles in the bottom panel of Fig.~\ref{fig:logJM}. Increasing the effective viscosity means less differential rotation, and thus the upper limit moves closer to the yellow line in Fig.~\ref{fig:logJM}. If in the future the sample can be complimented with stars more massive than $\sim$6\Msun, it could be possible to place upper limits on the viscosity. For now, the data does not allow to make any distinction between the viscosities tested in this work. We conclude that no external torques are required to explain the data of \cite{Aerts2025-AM}.

\begin{figure}
    \centering
    \includegraphics[width=\linewidth]{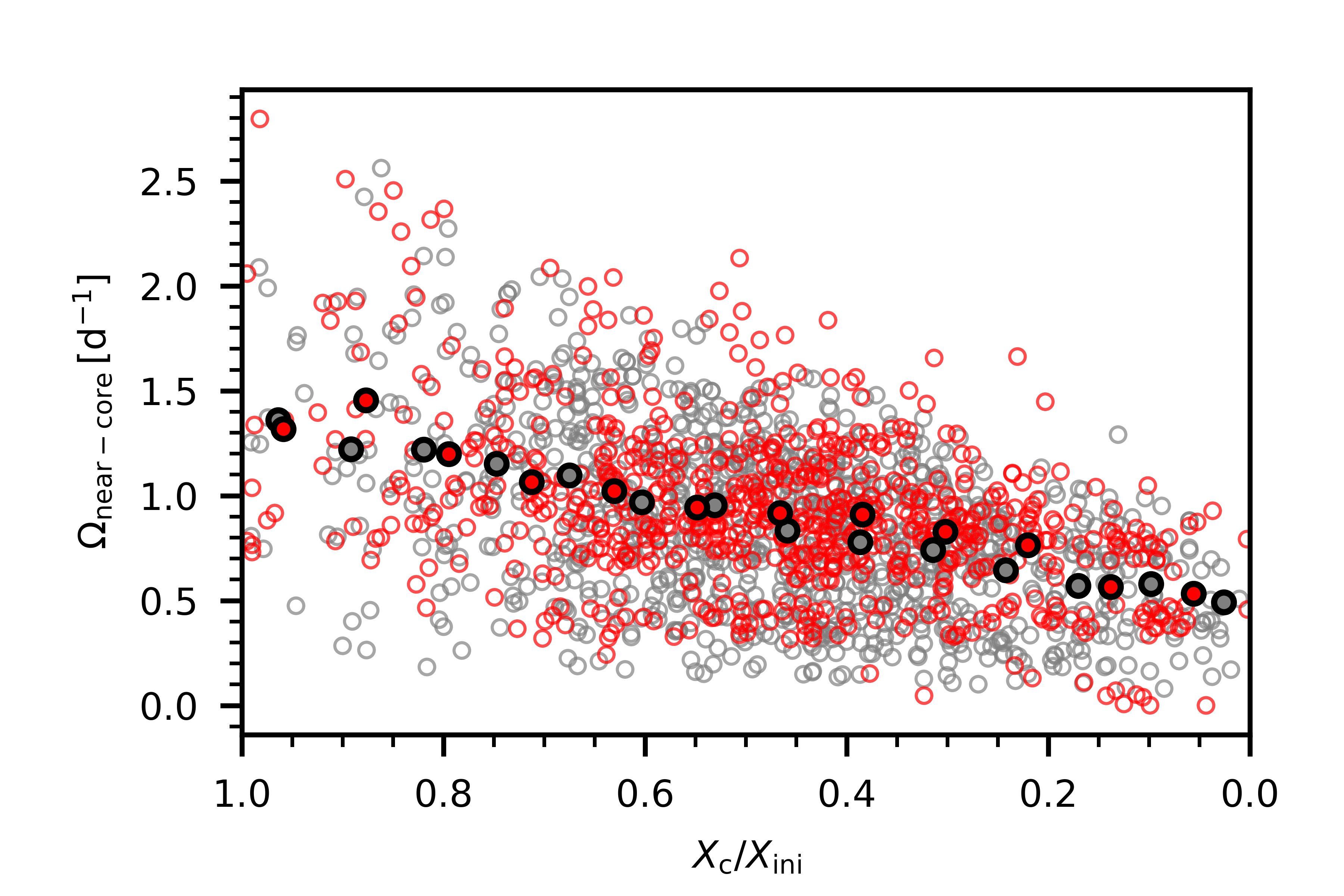}
    \caption{Near-core rotation frequencies versus normalised hydrogen-mass fraction in the core (age proxy). The red symbols show the sample of \cite{Aerts2025-AM}, the grey points show a simulated population of 1000 stars assuming a constant viscosity $\nu_{\rm eff} = 10^7\,{\rm cm^2\,{\rm s^{-1}}}$. The symbols with black edges indicate the bin averaged values, where the bin size is computed according to Scott's rule.  }
    \label{fig:XcX-Omega}
\end{figure}

\begin{figure}
    \centering
    \includegraphics[width=\linewidth]{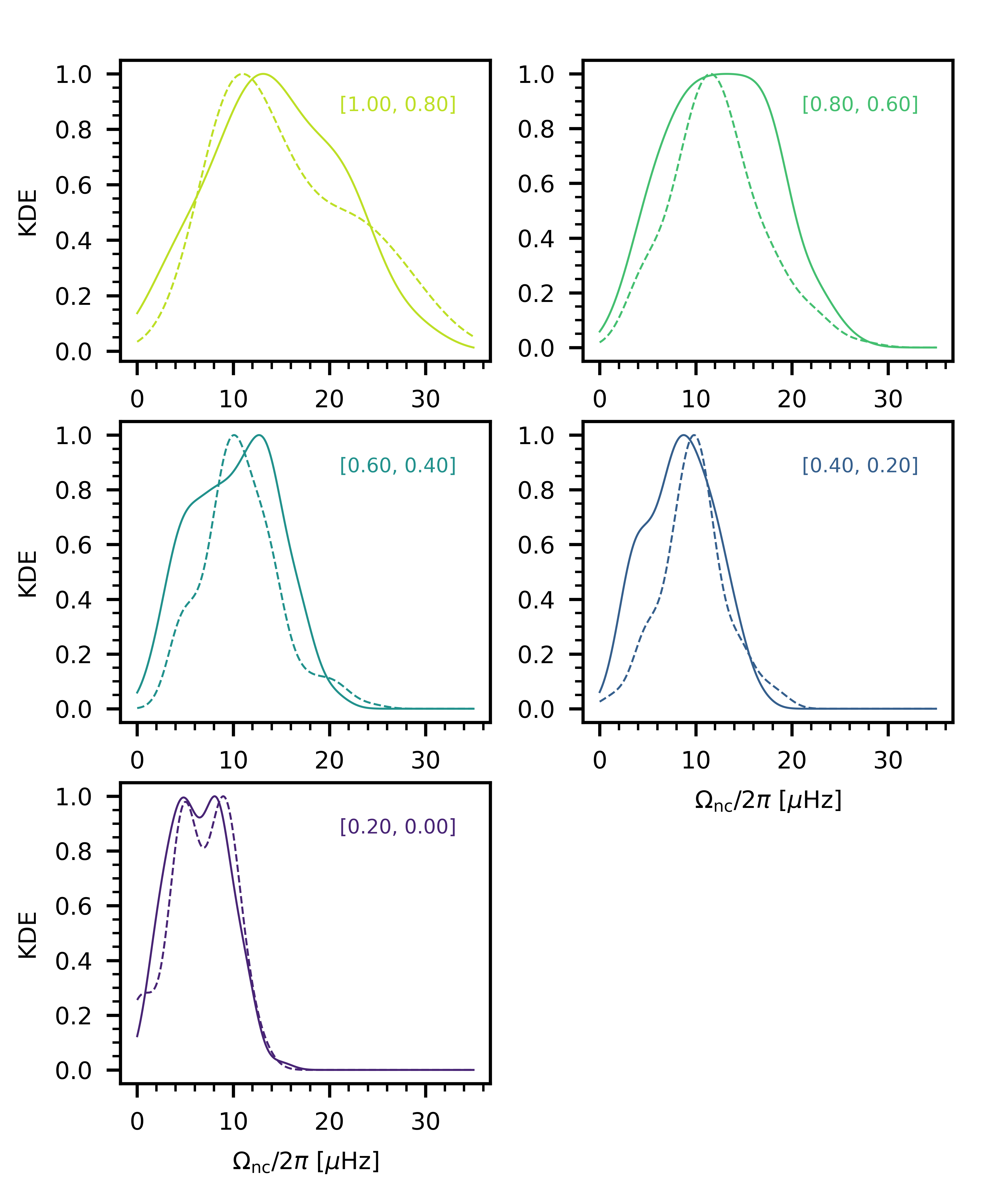}
    \caption{Distributions of the near-core rotation frequency for five \xcx bins across the main sequence. The dashed lines indicate the sample of \cite{Aerts2025-AM}, and the solid lines indicate the simulated population assuming a constant viscosity $\nu_{\rm eff} = 10^7\,{\rm cm^2\,{\rm s^{-1}}}$.}
    \label{fig:frot_distr_XcX}
\end{figure}

\begin{figure}
    \centering
    \includegraphics[width=\linewidth]{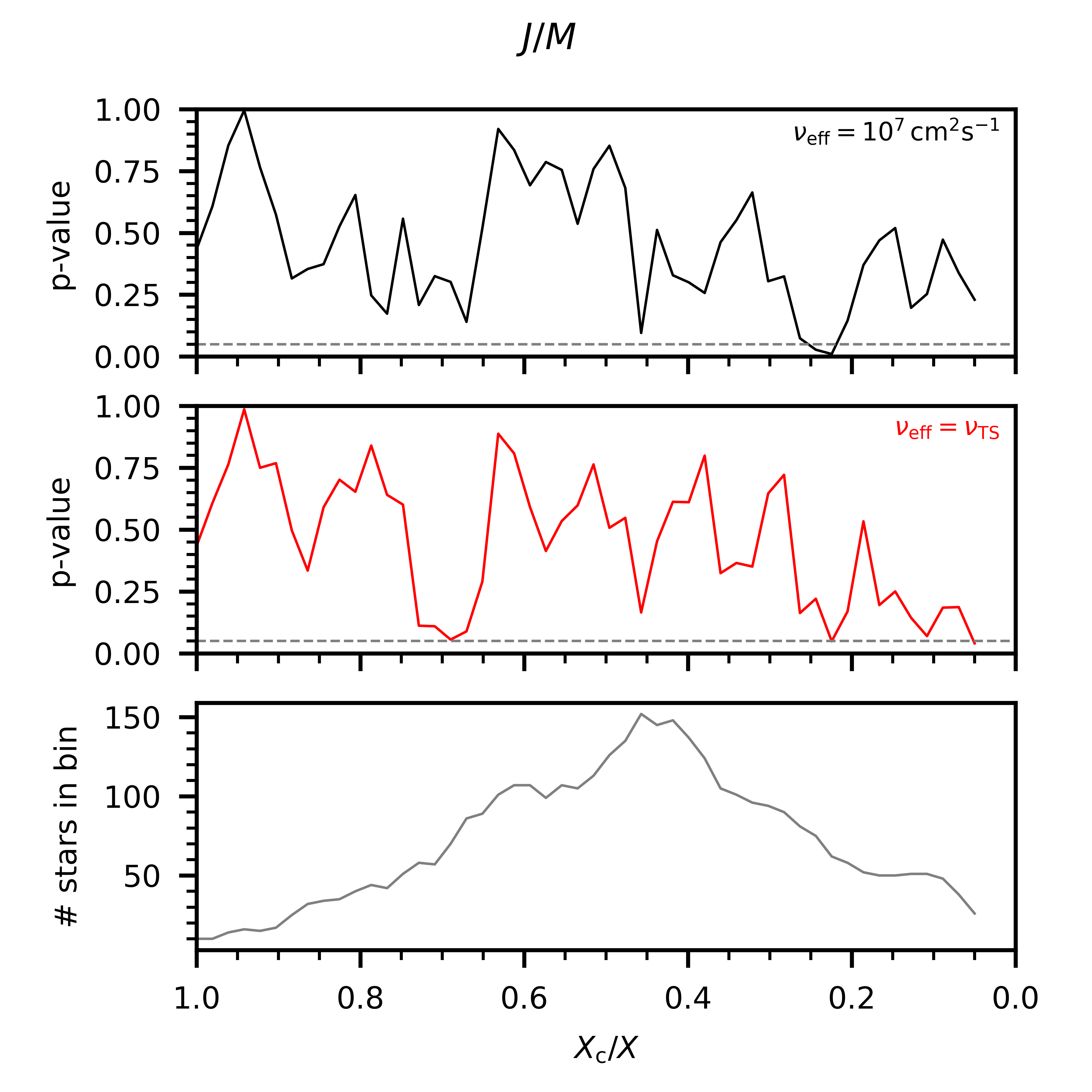}
    \caption{p-value of two-sample KS tests for the observed and simulated sample using a sliding \xcx bin of size 0.1. The upper panel corresponds to the grid with constant viscosity, the bottom panel to the grid with the TS dynamo. We reject the hypothesis that the observed and simulated $J/M$ distributions are different if the p-value is below 0.05, indicated by the horizontal dashed line.}
    \label{fig:p-value_frot}
\end{figure}

\section{Modelling time evolution of $\Omega$ and $J/M$}
We first start with the stars with masses $\ge 1.7$\Msun, of which there are 702 in the sample. We consider the grid with a constant viscosity $\nu_{\rm eff} = 10^7\,{\rm cm^2\,{\rm s^{-1}}}$. Sampling a population of 1000 stars from the observed mass and \xcx distributions, we find that a normal distribution of the initial rotation frequencies at the ZAMS with $(\overline{\omega_0},\sigma_{\omega_0}) = (0.3, 0.25)$ reproduces the observed $J/M$ values quite well. We truncate the normal distribution between $\overline{\omega_0} = 0.05$ and 0.55, given the limits of our grids. Figure~\ref{fig:XcX-Omega} shows the predicted distribution of near-core rotation frequencies across the main sequence in comparison with the observations from \cite{Aerts2025-AM}. We find a decrease of the average near-core rotation frequency per \xcx-bin that is similar to the observed one. In addition, the most extreme cases are well modelled too with our assumption of the initial rotation frequencies at the ZAMS. This is better visible in Fig.~\ref{fig:frot_distr_XcX}, where the kernel density estimates (KDEs) of the \frot distributions are shown for five sections along the main sequence.

To make a statistically underlined conclusion whether the observed and simulated PDFs are consistent, we perform two-sample Kolmogorov-Smirnov (KS) tests and assume the distributions are not drawn from the same underlying PDF if the p-value is lower than 0.05. For each age (\xcx) bin, we construct distributions of the rotation frequencies and specific AM. According to the p-values from the two-sample KS tests, the simulated distributions of the specific AM are consistent with observed ones across the main sequence, as shown in Fig.~\ref{fig:p-value_frot}. In other words, our priors on the mass, CBM, and initial rotation, as well as the assumed efficiency of the angular momentum transport are accurate enough to explain all available data on the 702 g-mode pulsators more massive than 1.7\Msun. The p-values that we find for the \frot distributions show that the simulated distributions are mostly consistent with the observed (model-independent) ones.

We find that decreasing or increasing the viscosity by one order of magnitude does not change the resulting \frot and $J/M$ distributions much. For the values of the effective viscosity tested here, only stars with masses $\gtrsim 7$\Msun and high initial rotation rates show differences in the near-core rotation frequency larger than the average observational uncertainty towards the very end of the main sequence (cf. Fig.~\ref{fig:omega_evol}). Such stars, however, are very uncommon in the sample. 

\begin{figure}
    \centering
    \centering
    \begin{subfigure}{0.49\textwidth}
        \centering
        \includegraphics[width=0.85\linewidth]{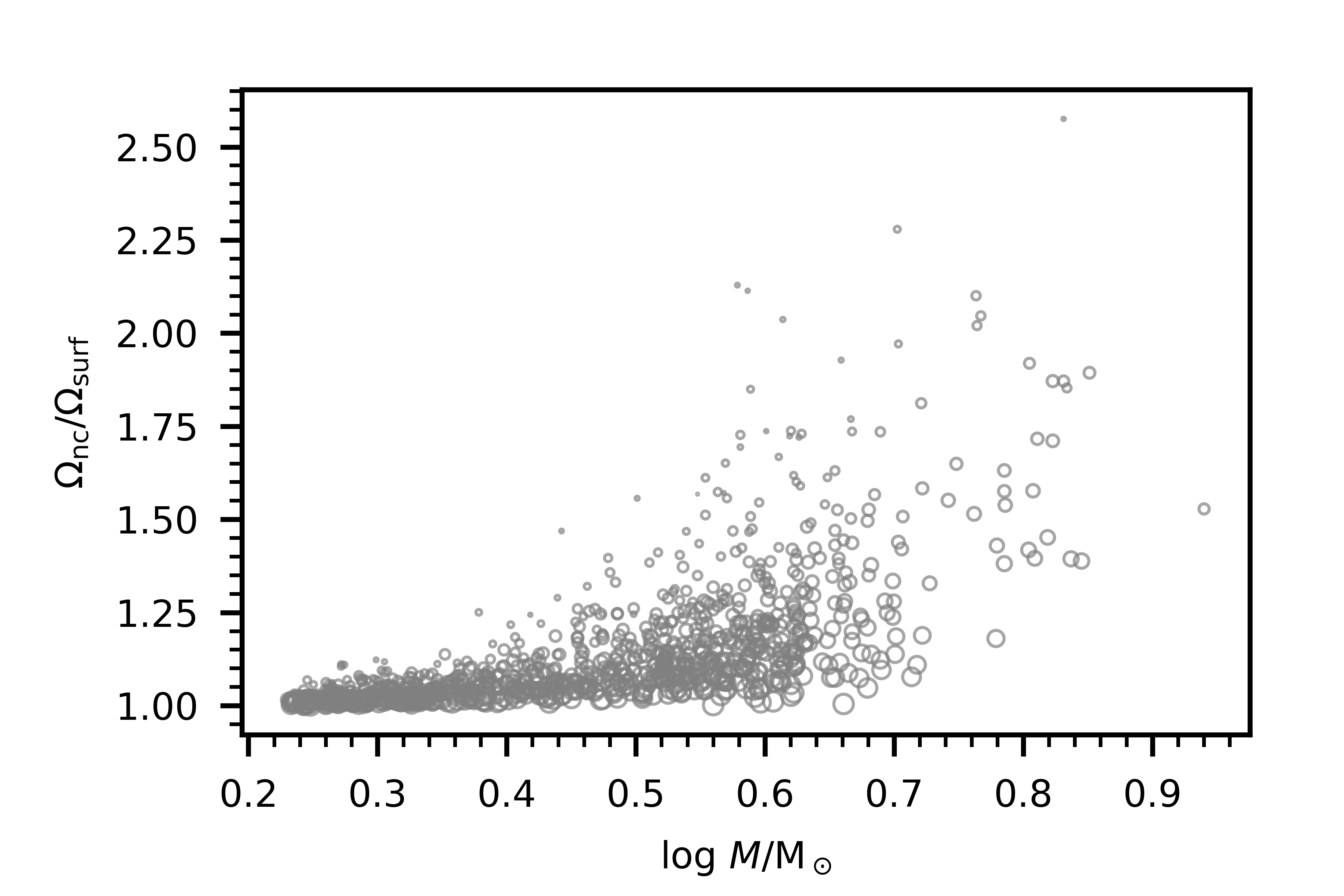}
        \caption{$\nu_{\rm eff} = 10^6\,{\rm cm^2\,{\rm s^{-1}}}$}
        \label{fig:sub1}
    \end{subfigure}
    \hfill
        \centering
    \begin{subfigure}{0.49\textwidth}
        \centering
        \includegraphics[width=0.85\linewidth]{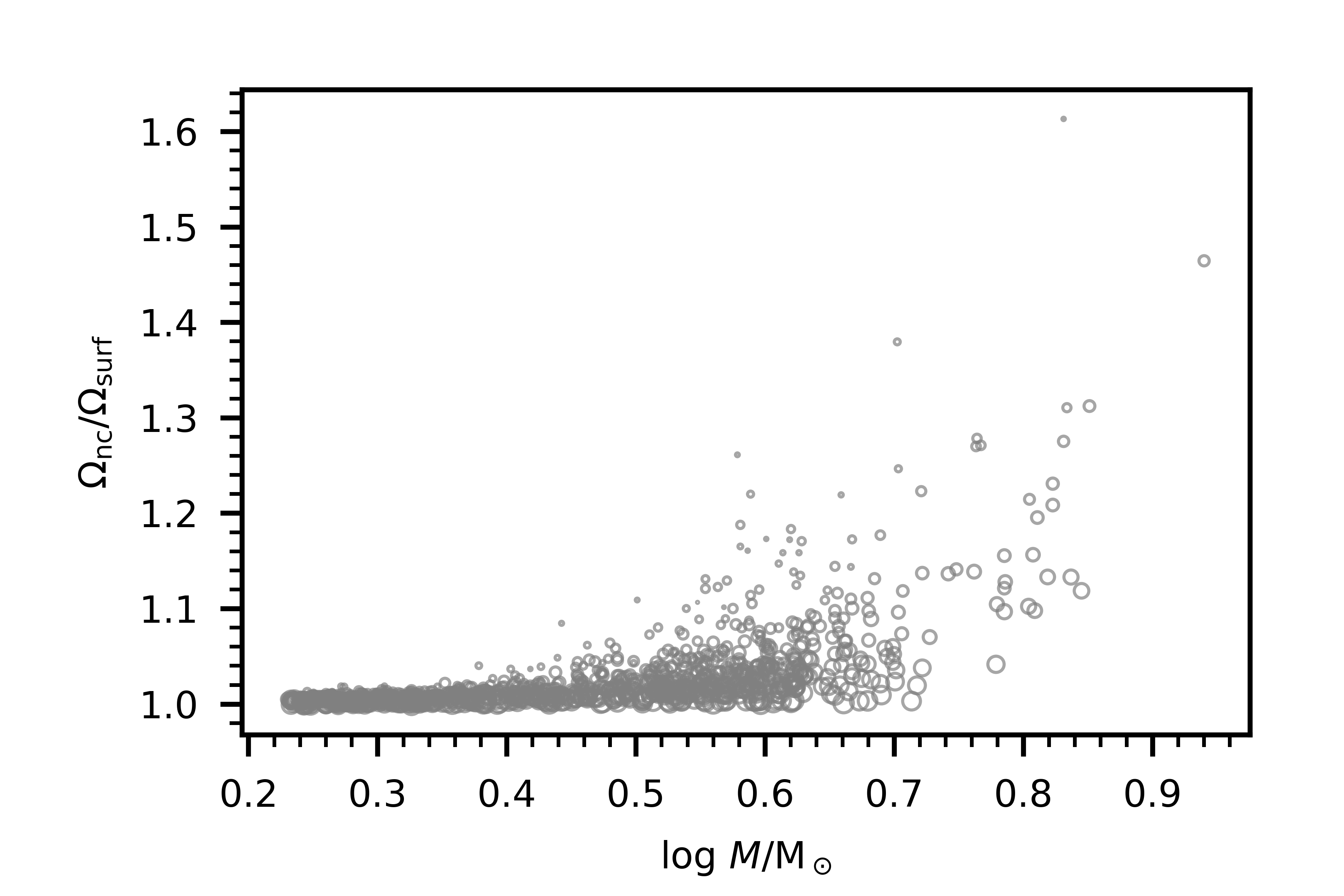}
        \caption{$\nu_{\rm eff} = 10^7\,{\rm cm^2\,{\rm s^{-1}}}$}
        \label{fig:sub1}
    \end{subfigure}
    \hfill
        \centering
    \begin{subfigure}{0.49\textwidth}
        \centering
        \includegraphics[width=0.85\linewidth]{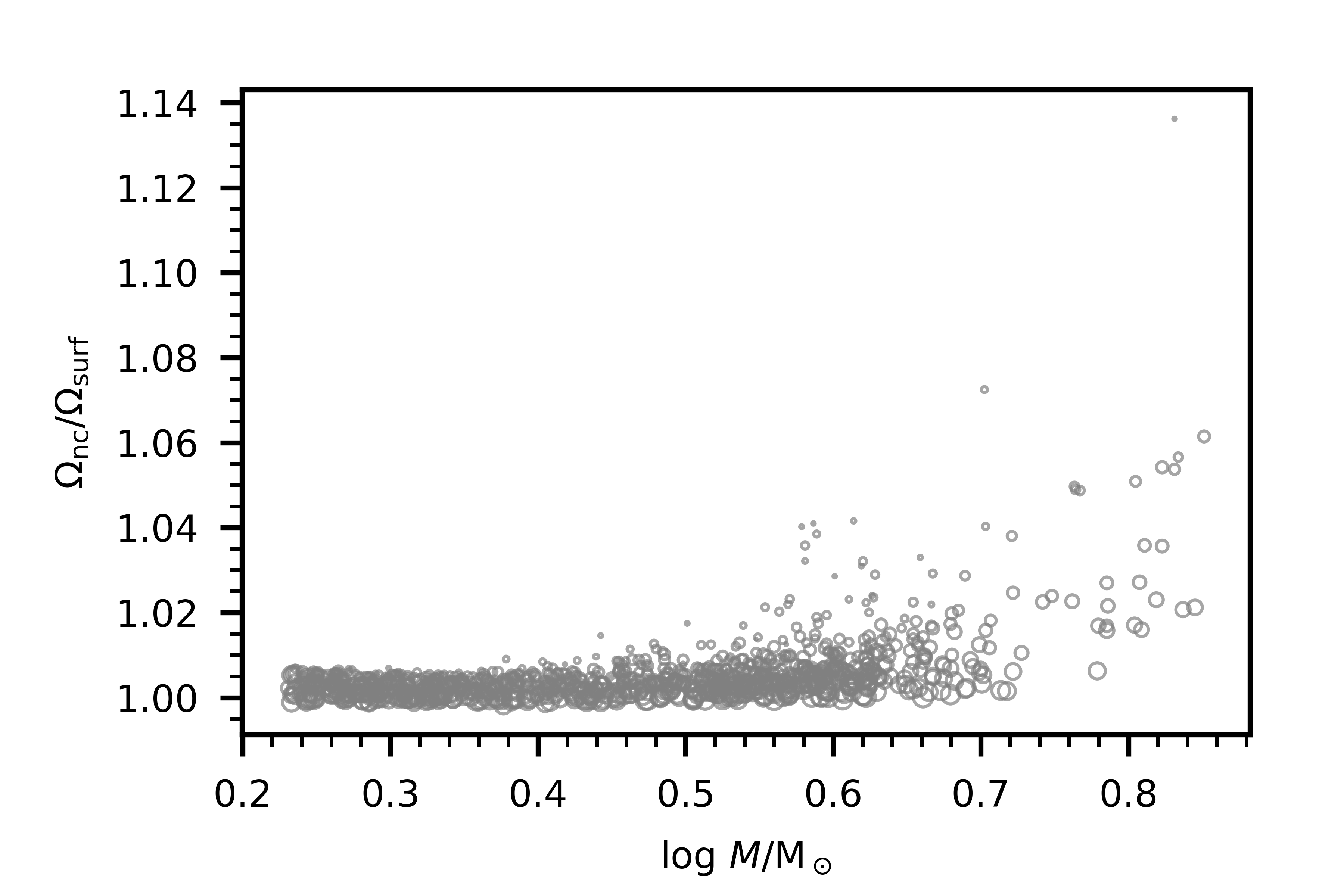}
        \caption{$\nu_{\rm eff} = 10^8\,{\rm cm^2\,{\rm s^{-1}}}$}
        \label{fig:sub1}
    \end{subfigure}
    \hfill
    \begin{subfigure}{0.49\textwidth}
        \centering
        \includegraphics[width=0.85\linewidth]{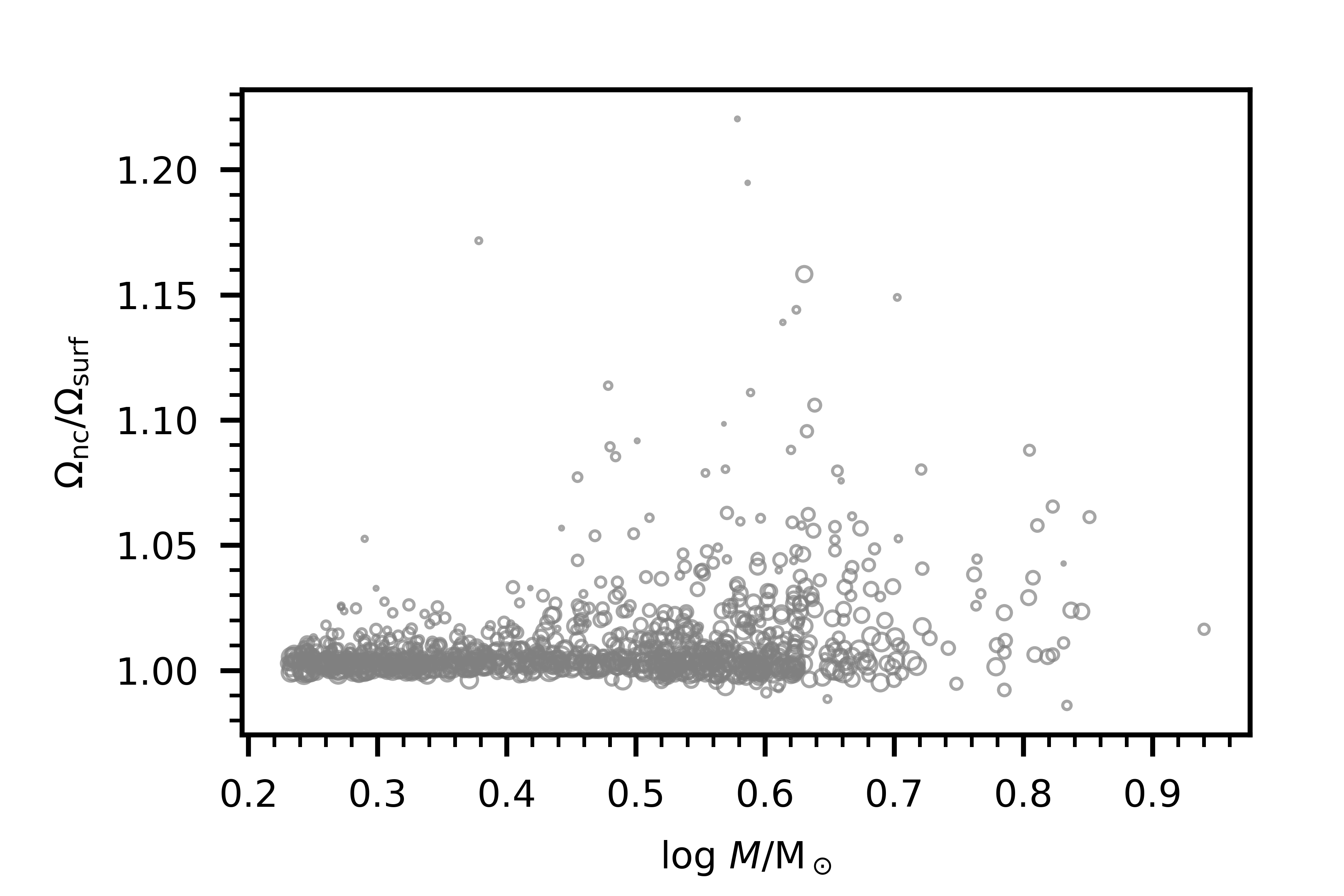}
        \caption{$\nu_{\rm eff} = \nu_{\rm TS}$}
        \label{fig:sub2}
    \end{subfigure}
    \caption{Ratio of the near-core rotation frequency to the average surface rotation frequency as a function of stellar mass for a simulated population of 1000 stars for different assumptions on the AM transport. The symbol size scales linearly with \xcx. }
    \label{fig:omegac-omegas}
\end{figure}

\section{Modelling differential rotation}
Although it is not easy to directly measure surface rotation frequencies of B-type stars with asteroseismology, we also make predictions for the level of radial differential rotation. Figure~\ref{fig:omegac-omegas} shows the predicted distribution of core-to-surface rotation rates as a function of mass for a population of 1000 stars. As can been seen in panel (b), for a constant $\nu_{\rm eff} = 10^7\,{\rm cm^2\,{\rm s^{-1}}}$, stars below roughly 3\Msun show very small levels of differential rotation and for masses below 9\Msun we expect most stars to have $\Omega_{\rm nc}/\Omega_{\rm surf} < 1.3$. A viscosity 10 times smaller (larger) yields differential rotation rates mostly below $\Omega_{\rm nc}/\Omega_{\rm surf} < 2$ (1.08). In case of the TS dynamo, the highest level of differential rotation in the simulated population is expected around 4\Msun and $\Omega_{\rm nc}/\Omega_{\rm surf} \lesssim 1.2$. In the models, however, the level of differential rotation that is developing as similar across mass (slightly lower for the lowest masses). Therefore, the peak in the simulated population is the result of the fact that the number of stars in the sample \cite{Aerts2025-AM} sharply decreases around 4\Msun. This highlights how observed measurements can be misinterpreted if they are only compared to theoretical upper and lower limits. We also note that the level of differential rotation in the models with the TS dynamo decreases with the rotation frequency, as is expected from the dependence of the predicted viscosity on the rotation frequency. Future studies of g-mode pulsators on the main sequence should focus on populating such diagrams in order to further improve the verification of the theoretically predicted viscosities linked to the TS dynamo. In any case, the majority of stars is still expected to be close to uniform rotation as shown in Fig.~\ref{fig:omegac-omegas}. In the case of a few \gdor stars, the ratio of the near-core rotation frequency to the rotation frequency of the convective core itself has been measured \citep[][not used in this study]{Saio2021}. \cite{Moyano2024} argue that, based on these ratios, the prescription of the TS dynamo of \cite{Spruit2002} is too efficient and that the associated viscosity needs to be scaled down by roughly three orders of magnitude. 
 
\begin{figure}
    \centering
    \includegraphics[width=\linewidth]{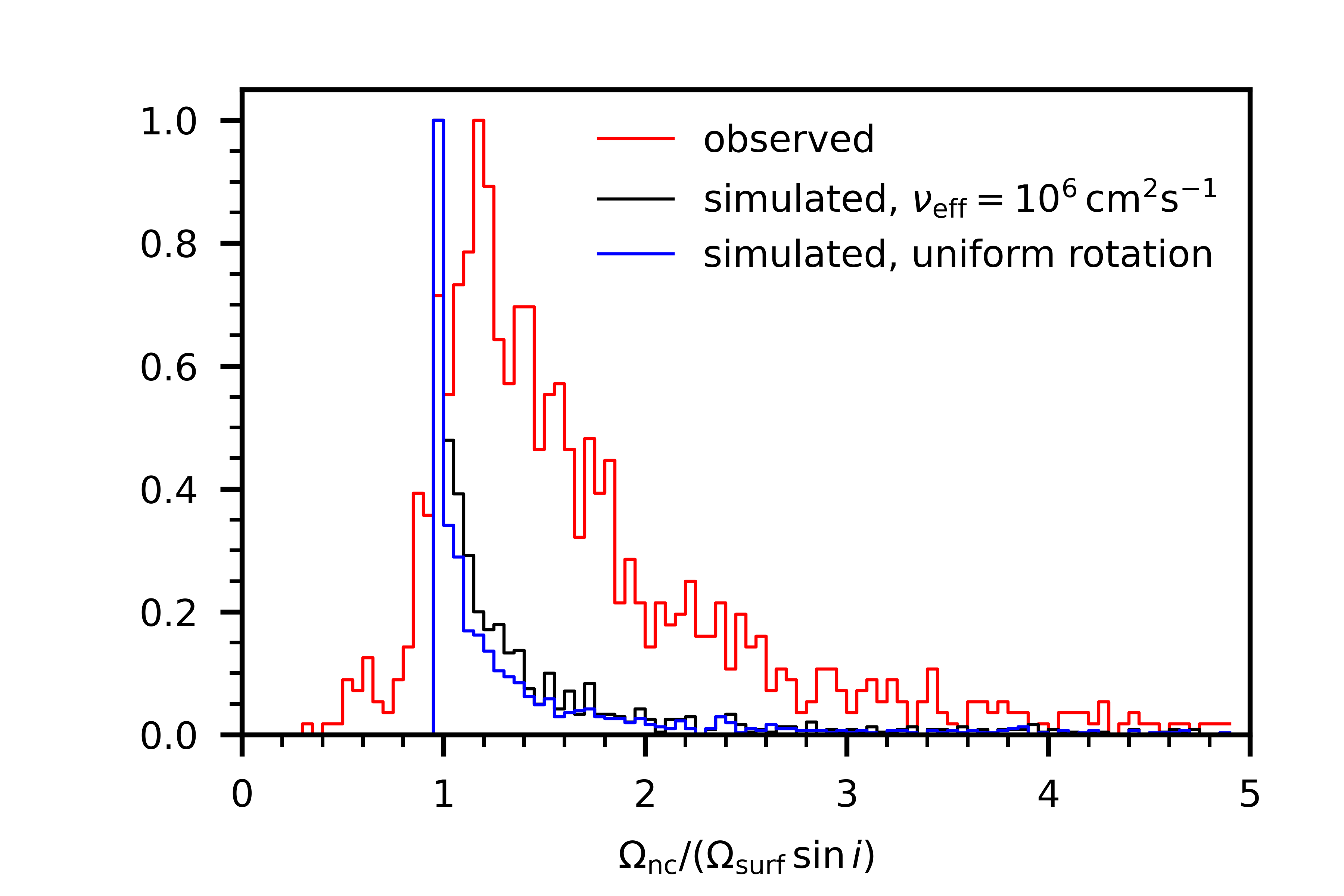}
    \caption{Observed and simulated distributions of the near-core rotation frequency to the projected surface rotation frequency. The assumed effective viscosity here is $\nu_{\rm eff} = 10^6\,{\rm cm^2\,{\rm s^{-1}}}$. The blue line shows the distribution that we expect when the stars are uniformly rotating, with a uniform distribution of the inclination. }
    \label{fig:fr_fs}
\end{figure}

Although no direct surface rotation frequencies are available, the sample of \cite{Aerts2025} does also contain 969 stars for which the projected surface rotation frequency $\Omega_{\rm surf}\,\sin i$ could be estimated from the Gaia \texttt{vbroad} measurements, as it provides an approximation of $ R_\star \Omega_{\rm surf}\cdot\sin i$, where the radius estimate comes from \cite{Mombarg2024-gaia}. We simulate another population of 1000 stars with masses between 1.4 and 4.5\Msun, corresponding to the masses covered in this subsample of \cite{Aerts2025}. For each star in the simulated population, we generate an inclination by uniformly sampling in $\cos i \sim U(0,1)$ and compute $\sin i = \sqrt{1 - \cos^2 i}$. We do note, however, that in principle prograde dipole modes are not expected to be observed close to $0^\circ$ due to cancellation effects. We find that our synthetic population mostly shows $\Omega_{\rm nc}/(\Omega_{\rm surf}\,\sin i)$ ratios close to one, while the sample of \cite{Aerts2025} peaks around 1.15, as shown in Fig.~\ref{fig:fr_fs}. The observations suggest that there is a small degree of differential rotation. In order to have the models develop this level of differential rotation, viscosities lower than $\nu_{\rm eff} = 10^6\,{\rm cm^2\,{\rm s^{-1}}}$ are required. There is some tension between the variance in the simulated population compared to the observed one. We note, however, that the measurement of $\Omega_{\rm nc}/(\Omega_{\rm surf}\,\sin i)$ has large uncertainties \citep{Aerts2025-AM} and that the (near-)core-to-surface rotation ratios derived by \cite{VanReeth2018}, \cite{Li2020}, and \cite{Saio2021} are close to one. 

Since the sample is dominated with lower-mass stars, varying the viscosity within the range that we study here, does not significantly change the resulting distribution. The distribution resulting from the TS grid is indistinguishable from uniform rotation. Furthermore, because the measurement of the projected surface rotation frequency comes from the observed rotational broadening, it is likely that this subsample of stars with measured surface rotation velocities is biased towards fast(er) rotating stars.

\section{Rotational chemical mixing}
\begin{figure}
    \centering
    \includegraphics[width=\linewidth]{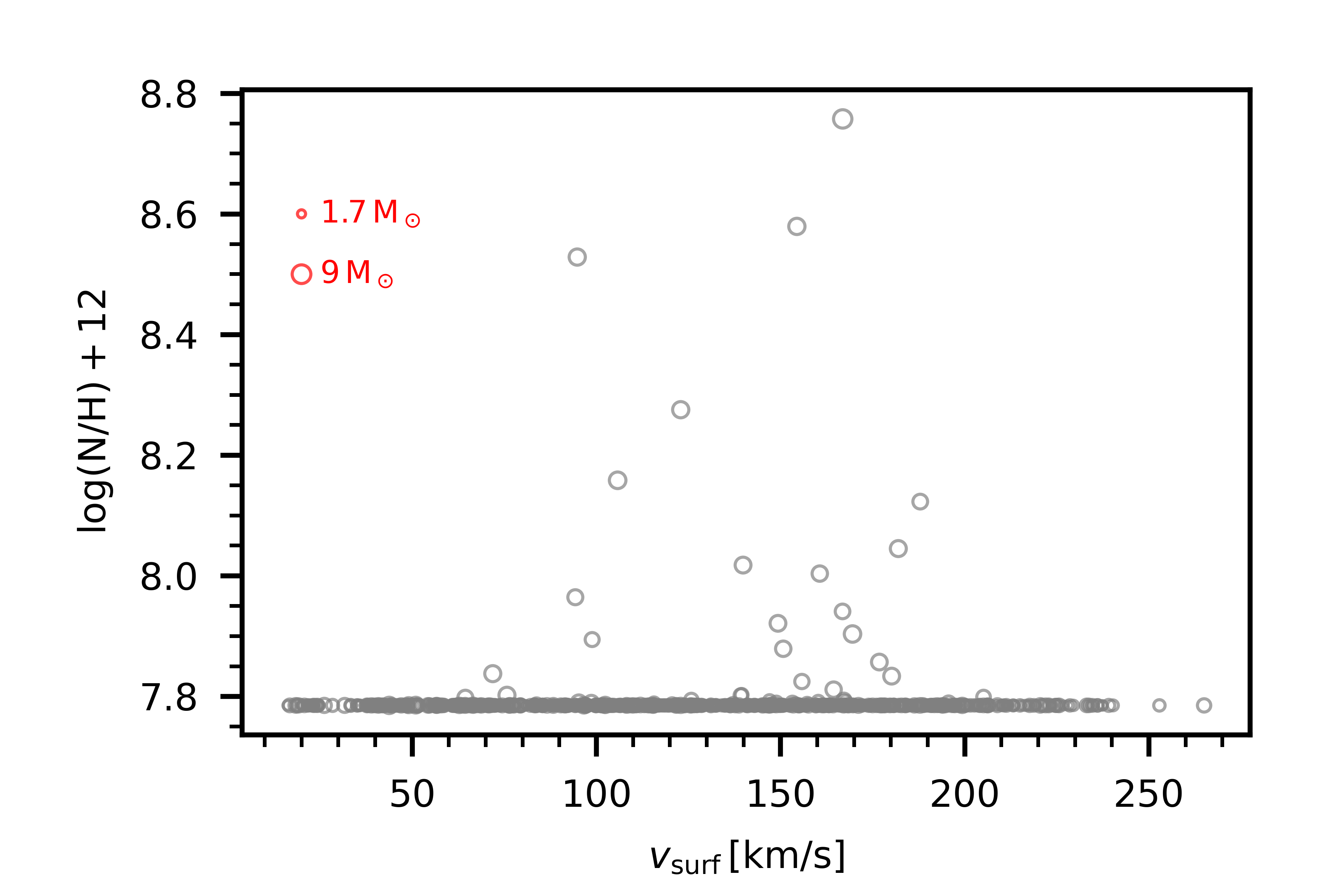}
    \caption{Simulated nitrogen-14 surface abundance as a function of the surface rotation velocity at the equator for a population of 1000 stars, assuming an effective viscosity of $\nu_{\rm eff} = 10^6\,{\rm cm^2\,{\rm s^{-1}}}$. The symbol sizes scale with the stellar mass. Most models do not show any enhancement at the surface compared to the baseline around 7.8 for $Z=0.014$ and a solar composition as per \cite{Asplund2009}. }
    \label{fig:n14}
\end{figure}

In this section, we have a look at the chemical mixing. If the chemical diffusion coefficient would be equal to the effective viscosities that we explore here, many of the models would experience chemically homogenous evolution. Therefore, we consider that the chemical diffusion coefficient is different from the effective viscosity, where the chemical mixing is driven by classical rotational mixing. We have applied a form of rotational chemical mixing in our stellar models following the works of \cite{Zahn1992} and \cite{ChaboyerZahn1992}, where the chemical diffusion coefficient scales as,

\begin{equation}
    D(r) = K \left(\frac{r}{N}  \frac{{\rm d} \Omega}{{\rm d}r} \right)^2,
\end{equation}
where $K$ is the thermal diffusivity, and $N$ is the Brunt-V\"ais\"al\"a frequency. We do not assume any minimal shear for this shear instability to be active.
Since the mixing efficiency scales with the gradient of the rotation profile, it is dependent on the efficiency of AM transport. The nitrogen-14 isotope is a useful tracer of the internal mixing as massive stars build up a nitrogen-14 excess in the convective core as a result of the CNO-cycle. If the mixing is efficient enough, the abundance at the surface can be enriched with material from the core. For an effective viscosity of $\nu_{\rm eff} = 10^6\,{\rm cm^2\,{\rm s^{-1}}}$, we predict that a small fraction of the (more massive) stars in the sample should have a measurable overabundance of nitrogen-14 at the surface, as shown in Fig.~\ref{fig:n14}. The range covered is consistent with spectroscopic measurements of SPBs carried out by \cite{Gebruers2021}, albeit that their sample size is rather small. The higher values of the effective viscosity do not result in any changes to the nitrogen-14 surface abundance throughout the main sequence. 

\section{Conclusion}
We have used a sample of 2937 g-mode pulsators from \cite{Aerts2025-AM} with characterized masses and ages \citep{Mombarg2024-gaia} and near-core rotation frequencies \citep{Aerts2025}, covering masses between 1.3 to 8.8\Msun, to place constraints on the efficiency of AM transport on the main sequence. Assuming AM is transported via diffusive processes only, where the efficiency is given by a constant effective viscosity, we find that viscosities on the order of $\sim 10^6\,{\rm cm^2\,{\rm s^{-1}}}$ can explain the observed evolution of the near-core rotation frequencies of g-mode pulsators. Additional models with viscosities predicted by the TS dynamo show that they meet this requirement of sufficiently large viscosity. Using the assumption of uniform rotation at the ZAMS, we can explain the data with initial critical rotation rates that follow a normal distribution, where there is a dichotomy between stars above and below 1.7\Msun. Future observational studies of main sequence g-mode pulsators should focus on measuring core-to-surface rotation rates (or rotation profiles via inversion techniques), particularly of B-type stars, as in this work we have fully exploited the constraining potential of near-core rotation frequencies alone. From the theoretical point-of-view, stellar models including different parametrisations for the transport of AM and chemical mixing have to be computed \citep[e.g.][]{Talon2005, fuller2019}. The predicted flux of AM must then be compared to the one corresponding to the value of the effective viscosity that we calibrate in this paper. This would allow us to rule out or confirm the possible action of the investigated mechanism.

\section*{Data availability}
The grids of \mesa models are available on Zenodo at: \url{https://zenodo.org/records/20731921}

\begin{acknowledgements}
We thank the anonymous referee for the remarks on the manuscript. 
  We also thank Conny Aerts for the useful discussions. The research leading to these results has received funding from the European Research Council (ERC) under the Horizon Europe programme (Synergy Grant agreement N$^\circ$101071505: 4D-STAR). While partially funded by the European Union, views and opinions expressed are however those of the authors only and do not necessarily reflect those of the European Union or the European Research Council. Neither the European Union nor the granting authority can be held responsible for them. The computational resources and services used in this work were provided by the VSC (Flemish Supercomputer Center),
  funded by the Research Foundation - Flanders (FWO) and the Flemish Government department EWI. S.M. acknowledges support from the PLATO CNES grant at CEA-irfu/DAp. This research made use of the \texttt{numpy} \citep{Harris2020} and \texttt{matplotlib} \citep{Hunter2007} \texttt{Python} software packages.  
\end{acknowledgements}

%
%
\bibliographystyle{aa} 
\bibliography{main} 

@ARTICLE{Paxton2011,
   author = {{Paxton}, B. and {Bildsten}, L. and {Dotter}, A. and {Herwig}, F. and 
	{Lesaffre}, P. and {Timmes}, F.},
    title = "{Modules for Experiments in Stellar Astrophysics (MESA)}",
  journal = {\apjs},
archivePrefix = "arXiv",
   eprint = {1009.1622},
 primaryClass = "astro-ph.SR",
     year = 2011,
    month = jan,
   volume = 192,
      eid = {3},
    pages = {3},
      doi = {10.1088/0067-0049/192/1/3},
   adsurl = {http://adsabs.harvard.edu/abs/2011ApJS..192....3P}
}

@ARTICLE{Paxton2013,
   author = {{Paxton}, B. and {Cantiello}, M. and {Arras}, P. and {Bildsten}, L. and 
	{Brown}, E.~F. and {Dotter}, A. and {Mankovich}, C. and {Montgomery}, M.~H. and 
	{Stello}, D. and {Timmes}, F.~X. and {Townsend}, R.},
    title = "{Modules for Experiments in Stellar Astrophysics (MESA): Planets, Oscillations, Rotation, and Massive Stars}",
  journal = {\apjs},
archivePrefix = "arXiv",
   eprint = {1301.0319},
 primaryClass = "astro-ph.SR",
     year = 2013,
    month = sep,
   volume = 208,
      eid = {4},
    pages = {4},
      doi = {10.1088/0067-0049/208/1/4},
   adsurl = {http://adsabs.harvard.edu/abs/2013ApJS..208....4P}
}

@ARTICLE{Paxton2015,
   author = {{Paxton}, B. and {Marchant}, P. and {Schwab}, J. and {Bauer}, E.~B. and 
	{Bildsten}, L. and {Cantiello}, M. and {Dessart}, L. and {Farmer}, R. and 
	{Hu}, H. and {Langer}, N. and {Townsend}, R.~H.~D. and {Townsley}, D.~M. and 
	{Timmes}, F.~X.},
    title = "{Modules for Experiments in Stellar Astrophysics (MESA): Binaries, Pulsations, and Explosions}",
  journal = {\apjs},
archivePrefix = "arXiv",
   eprint = {1506.03146},
 primaryClass = "astro-ph.SR",
     year = 2015,
    month = sep,
   volume = 220,
      eid = {15},
    pages = {15},
      doi = {10.1088/0067-0049/220/1/15},
   adsurl = {http://adsabs.harvard.edu/abs/2015ApJS..220...15P}
}

@ARTICLE{Paxton2018,
   author = {{Paxton}, B. and {Schwab}, J. and {Bauer}, E.~B. and {Bildsten}, L. and 
	{Blinnikov}, S. and {Duffell}, P. and {Farmer}, R. and {Goldberg}, J.~A. and 
	{Marchant}, P. and {Sorokina}, E. and {Thoul}, A. and {Townsend}, R.~H.~D. and 
	{Timmes}, F.~X.},
    title = "{Modules for Experiments in Stellar Astrophysics (MESA): Convective Boundaries, Element Diffusion, and Massive Star Explosions}",
  journal = {\apjs},
archivePrefix = "arXiv",
   eprint = {1710.08424},
 primaryClass = "astro-ph.SR",
     year = 2018,
    month = feb,
   volume = 234,
      eid = {34},
    pages = {34},
      doi = {10.3847/1538-4365/aaa5a8},
   adsurl = {http://adsabs.harvard.edu/abs/2018ApJS..234...34P}
}

@ARTICLE{Asplund2009,
   author = {{Asplund}, M. and {Grevesse}, N. and {Sauval}, A.~J. and {Scott}, P.
	},
    title = "{The Chemical Composition of the Sun}",
  journal = {\araa},
archivePrefix = "arXiv",
   eprint = {0909.0948},
 primaryClass = "astro-ph.SR",
     year = 2009,
    month = sep,
   volume = 47,
    pages = {481-522},
      doi = {10.1146/annurev.astro.46.060407.145222},
   adsurl = {http://adsabs.harvard.edu/abs/2009ARA%26A..47..481A}
}

@ARTICLE{Aerts2019-ARAA,
       author = {{Aerts}, Conny and {Mathis}, St{\'e}phane and {Rogers}, Tamara M.},
        title = "{Angular Momentum Transport in Stellar Interiors}",
      journal = {\araa},
         year = "2019",
        month = "Aug",
       volume = {57},
        pages = {35-78},
          doi = {10.1146/annurev-astro-091918-104359},
archivePrefix = {arXiv},
       eprint = {1809.07779},
 primaryClass = {astro-ph.SR},
       adsurl = {https://ui.adsabs.harvard.edu/abs/2019ARA&A..57...35A}
}

@ARTICLE{Paxton2019,
       author = {{Paxton}, Bill and {Smolec}, R. and {Schwab}, Josiah and {Gautschy}, A. and
         {Bildsten}, Lars and {Cantiello}, Matteo and {Dotter}, Aaron and
         {Farmer}, R. and {Goldberg}, Jared A. and {Jermyn}, Adam S. and
         {Kanbur}, S.~M. and {Marchant}, Pablo and {Thoul}, Anne and
         {Townsend}, Richard H.~D. and {Wolf}, William M. and {Zhang}, Michael and
         {Timmes}, F.~X.},
        title = "{Modules for Experiments in Stellar Astrophysics (MESA): Pulsating Variable Stars, Rotation, Convective Boundaries, and Energy Conservation}",
      journal = {\apjs},
         year = "2019",
        month = "Jul",
       volume = {243},
       number = {1},
          eid = {10},
        pages = {10},
          doi = {10.3847/1538-4365/ab2241},
archivePrefix = {arXiv},
       eprint = {1903.01426},
 primaryClass = {astro-ph.SR},
       adsurl = {https://ui.adsabs.harvard.edu/abs/2019ApJS..243...10P}
}

@ARTICLE{VanReeth2015b,
   author = {{Van Reeth}, T. and {Tkachenko}, A. and {Aerts}, C. and {P{\'a}pics}, P.~I. and 
	{Triana}, S.~A. and {Zwintz}, K. and {Degroote}, P. and {Debosscher}, J. and 
	{Bloemen}, S. and {Schmid}, V.~S. and {De Smedt}, K. and {Fremat}, Y. and 
	{Fuentes}, A.~S. and {Homan}, W. and {Hrudkova}, M. and {Karjalainen}, R. and 
	{Lombaert}, R. and {Nemeth}, P. and {{\O}stensen}, R. and {Van De Steene}, G. and 
	{Vos}, J. and {Raskin}, G. and {Van Winckel}, H.},
    title = "{Gravity-mode Period Spacings as a Seismic Diagnostic for a Sample of {$\gamma$} Doradus Stars from Kepler Space Photometry and High-resolution Ground-based Spectroscopy}",
  journal = {\apjs},
archivePrefix = "arXiv",
   eprint = {1504.02119},
 primaryClass = "astro-ph.SR",
     year = 2015,
    month = jun,
   volume = 218,
      eid = {27},
    pages = {27},
      doi = {10.1088/0067-0049/218/2/27},
   adsurl = {http://adsabs.harvard.edu/abs/2015ApJS..218...27V}
}

@ARTICLE{VanReeth2018,
   author = {{Van Reeth}, T. and {Mombarg}, J.~S.~G. and {Mathis}, S. and 
	{Tkachenko}, A. and {Fuller}, J. and {Bowman}, D.~M. and {Buysschaert}, B. and 
	{Johnston}, C. and {Garc{\'{\i}}a Hern{\'a}ndez}, A. and {Goldstein}, J. and 
	{Townsend}, R.~H.~D. and {Aerts}, C.},
    title = "{Sensitivity of gravito-inertial modes to differential rotation in intermediate-mass main-sequence stars}",
  journal = {\aap},
archivePrefix = "arXiv",
   eprint = {1806.03586},
 primaryClass = "astro-ph.SR",
     year = 2018,
    month = oct,
   volume = 618,
      eid = {A24},
    pages = {A24},
      doi = {10.1051/0004-6361/201832718},
   adsurl = {http://adsabs.harvard.edu/abs/2018A%26A...618A..24V}
}

@ARTICLE{Ouazzani2019,
       author = {{Ouazzani}, R. -M. and {Marques}, J.~P. and {Goupil}, M. -J. and
         {Christophe}, S. and {Antoci}, V. and {Salmon}, S.~J.~A.~J. and
         {Ballot}, J.},
        title = "{{\ensuremath{\gamma}} Doradus stars as a test of angular momentum transport models}",
      journal = {\aap},
         year = "2019",
        month = "Jun",
       volume = {626},
          eid = {A121},
        pages = {A121},
          doi = {10.1051/0004-6361/201832607},
archivePrefix = {arXiv},
       eprint = {1801.09228},
 primaryClass = {astro-ph.SR},
       adsurl = {https://ui.adsabs.harvard.edu/abs/2019A&A...626A.121O}
}

@ARTICLE{Freytag1996,
       author = {{Freytag}, B. and {Ludwig}, H. -G. and {Steffen}, M.},
        title = "{Hydrodynamical models of stellar convection. The role of overshoot in DA white dwarfs, A-type stars, and the Sun.}",
      journal = {\aap},
         year = "1996",
        month = "Sep",
       volume = {313},
        pages = {497-516},
       adsurl = {https://ui.adsabs.harvard.edu/abs/1996A&A...313..497F}
}

@ARTICLE{fuller2019,
       author = {{Fuller}, Jim and {Piro}, Anthony L. and {Jermyn}, Adam S.},
        title = "{Slowing the spins of stellar cores}",
      journal = {\mnras},
         year = 2019,
        month = may,
       volume = {485},
       number = {3},
        pages = {3661-3680},
          doi = {10.1093/mnras/stz514},
archivePrefix = {arXiv},
       eprint = {1902.08227},
 primaryClass = {astro-ph.SR},
       adsurl = {https://ui.adsabs.harvard.edu/abs/2019MNRAS.485.3661F}
}

@ARTICLE{Li2020,
       author = {{Li}, Gang and {Van Reeth}, Timothy and {Bedding}, Timothy R. and
         {Murphy}, Simon J. and {Antoci}, Victoria and {Ouazzani}, Rhita-Maria and
         {Barbara}, Nicholas H.},
        title = "{Gravity-mode period spacings and near-core rotation rates of 611 {\ensuremath{\gamma}} Doradus stars with Kepler}",
      journal = {\mnras},
         year = 2020,
        month = jan,
       volume = {491},
       number = {3},
        pages = {3586-3605},
          doi = {10.1093/mnras/stz2906},
archivePrefix = {arXiv},
       eprint = {1910.06634},
 primaryClass = {astro-ph.SR},
       adsurl = {https://ui.adsabs.harvard.edu/abs/2020MNRAS.491.3586L}
}

@ARTICLE{dupret2004,
       author = {{Dupret}, M. -A. and {Grigahc{\`e}ne}, A. and {Garrido}, R. and {Gabriel}, M. and {Scuflaire}, R.},
        title = "{Theoretical instability strips for {\ensuremath{\delta}} Scuti and {\ensuremath{\gamma}} Doradus stars}",
      journal = {\aap},
         year = 2004,
        month = jan,
       volume = {414},
        pages = {L17-L20},
          doi = {10.1051/0004-6361:20031740},
       adsurl = {https://ui.adsabs.harvard.edu/abs/2004A&A...414L..17D}
}

@ARTICLE{Saio2021,
       author = {{Saio}, Hideyuki and {Takata}, Masao and {Lee}, Umin and {Li}, Gang and {Van Reeth}, Timothy},
        title = "{Rotation of the convective core in {\ensuremath{\gamma}} Dor stars measured by dips in period spacings of g modes coupled with inertial modes}",
      journal = {\mnras},
         year = 2021,
        month = apr,
       volume = {502},
       number = {4},
        pages = {5856-5874},
          doi = {10.1093/mnras/stab482},
archivePrefix = {arXiv},
       eprint = {2102.08548},
 primaryClass = {astro-ph.SR},
       adsurl = {https://ui.adsabs.harvard.edu/abs/2021MNRAS.502.5856S}
}

@ARTICLE{Pedersen2021,
       author = {{Pedersen}, May G. and {Aerts}, Conny and {P{\'a}pics}, P{\'e}ter I. and {Michielsen}, Mathias and {Gebruers}, Sarah and {Rogers}, Tamara M. and {Molenberghs}, Geert and {Burssens}, Siemen and {Garcia}, Stefano and {Bowman}, Dominic M.},
        title = "{Internal mixing of rotating stars inferred from dipole gravity modes}",
      journal = {Nature Astronomy},
         year = 2021,
        month = jan,
       volume = {5},
        pages = {715-722},
          doi = {10.1038/s41550-021-01351-x},
archivePrefix = {arXiv},
       eprint = {2105.04533},
 primaryClass = {astro-ph.SR},
       adsurl = {https://ui.adsabs.harvard.edu/abs/2021NatAs...5..715P}
}

@ARTICLE{Heger2000,
       author = {{Heger}, A. and {Langer}, N. and {Woosley}, S.~E.},
        title = "{Presupernova Evolution of Rotating Massive Stars. I. Numerical Method and Evolution of the Internal Stellar Structure}",
      journal = {\apj},
         year = 2000,
        month = jan,
       volume = {528},
       number = {1},
        pages = {368-396},
          doi = {10.1086/308158},
archivePrefix = {arXiv},
       eprint = {astro-ph/9904132},
 primaryClass = {astro-ph},
       adsurl = {https://ui.adsabs.harvard.edu/abs/2000ApJ...528..368H}
}

@ARTICLE{Gebruers2021,
       author = {{Gebruers}, Sarah and {Straumit}, Ilya and {Tkachenko}, Andrew and {Mombarg}, Joey S.~G. and {Pedersen}, May G. and {Van Reeth}, Timothy and {Li}, Gang and {Lampens}, Patricia and {Escorza}, Ana and {Bowman}, Dominic M. and {De Cat}, Peter and {Vermeylen}, Lore and {Bodensteiner}, Julia and {Rix}, Hans-Walter and {Aerts}, Conny},
        title = "{A homogeneous spectroscopic analysis of a Kepler legacy sample of dwarfs for gravity-mode asteroseismology}",
      journal = {\aap},
         year = 2021,
        month = jun,
       volume = {650},
          eid = {A151},
        pages = {A151},
          doi = {10.1051/0004-6361/202140466},
archivePrefix = {arXiv},
       eprint = {2104.04521},
 primaryClass = {astro-ph.SR},
       adsurl = {https://ui.adsabs.harvard.edu/abs/2021A&A...650A.151G}
}

@ARTICLE{Pedersen2022b,
       author = {{Pedersen}, May G.},
        title = "{Internal Rotation and Inclinations of Slowly Pulsating B Stars: Evidence of Interior Angular Momentum Transport}",
      journal = {\apj},
         year = 2022,
        month = nov,
       volume = {940},
       number = {1},
          eid = {49},
        pages = {49},
          doi = {10.3847/1538-4357/ac947f},
archivePrefix = {arXiv},
       eprint = {2208.14497},
 primaryClass = {astro-ph.SR},
       adsurl = {https://ui.adsabs.harvard.edu/abs/2022ApJ...940...49P}
}

@ARTICLE{Zahn1992,
       author = {{Zahn}, J. -P.},
        title = "{Circulation and turbulence in rotating stars.}",
      journal = {\aap},
         year = 1992,
        month = nov,
       volume = {265},
        pages = {115-132},
       adsurl = {https://ui.adsabs.harvard.edu/abs/1992A&A...265..115Z}
}

@Article{Harris2020,
 title         = {Array programming with {NumPy}},
 author        = {Charles R. Harris and K. Jarrod Millman and St{\'{e}}fan J.
                 van der Walt and Ralf Gommers and Pauli Virtanen and David
                 Cournapeau and Eric Wieser and Julian Taylor and Sebastian
                 Berg and Nathaniel J. Smith and Robert Kern and Matti Picus
                 and Stephan Hoyer and Marten H. van Kerkwijk and Matthew
                 Brett and Allan Haldane and Jaime Fern{\'{a}}ndez del
                 R{\'{i}}o and Mark Wiebe and Pearu Peterson and Pierre
                 G{\'{e}}rard-Marchant and Kevin Sheppard and Tyler Reddy and
                 Warren Weckesser and Hameer Abbasi and Christoph Gohlke and
                 Travis E. Oliphant},
 year          = {2020},
 month         = sep,
 journal       = {Nature},
 volume        = {585},
 number        = {7825},
 pages         = {357--362},
 doi           = {10.1038/s41586-020-2649-2},
 publisher     = {Springer Science and Business Media {LLC}},
 url           = {https://doi.org/10.1038/s41586-020-2649-2}
}

@Article{Hunter2007,
  Author    = {Hunter, J. D.},
  Title     = {Matplotlib: A 2D graphics environment},
  Journal   = {Computing in Science \& Engineering},
  Volume    = {9},
  Number    = {3},
  Pages     = {90--95},
  publisher = {IEEE COMPUTER SOC},
  doi       = {10.1109/MCSE.2007.55},
  year      = 2007
}

@ARTICLE{Spruit2002,
       author = {{Spruit}, H.~C.},
        title = "{Dynamo action by differential rotation in a stably stratified stellar interior}",
      journal = {\aap},
         year = 2002,
        month = jan,
       volume = {381},
        pages = {923-932},
          doi = {10.1051/0004-6361:20011465},
archivePrefix = {arXiv},
       eprint = {astro-ph/0108207},
 primaryClass = {astro-ph},
       adsurl = {https://ui.adsabs.harvard.edu/abs/2002A&A...381..923S}
}

@ARTICLE{Jermyn2023,
       author = {{Jermyn}, Adam S. and {Bauer}, Evan B. and {Schwab}, Josiah and {Farmer}, R. and {Ball}, Warrick H. and {Bellinger}, Earl P. and {Dotter}, Aaron and {Joyce}, Meridith and {Marchant}, Pablo and {Mombarg}, Joey S.~G. and {Wolf}, William M. and {Sunny Wong}, Tin Long and {Cinquegrana}, Giulia C. and {Farrell}, Eoin and {Smolec}, R. and {Thoul}, Anne and {Cantiello}, Matteo and {Herwig}, Falk and {Toloza}, Odette and {Bildsten}, Lars and {Townsend}, Richard H.~D. and {Timmes}, F.~X.},
        title = "{Modules for Experiments in Stellar Astrophysics (MESA): Time-dependent Convection, Energy Conservation, Automatic Differentiation, and Infrastructure}",
      journal = {\apjs},
         year = 2023,
        month = mar,
       volume = {265},
       number = {1},
          eid = {15},
        pages = {15},
          doi = {10.3847/1538-4365/acae8d},
archivePrefix = {arXiv},
       eprint = {2208.03651},
 primaryClass = {astro-ph.SR},
       adsurl = {https://ui.adsabs.harvard.edu/abs/2023ApJS..265...15J}
}

@ARTICLE{Talon2005,
       author = {{Talon}, S. and {Charbonnel}, C.},
        title = "{Hydrodynamical stellar models including rotation, internal gravity waves, and atomic diffusion. I. Formalism and tests on Pop I dwarfs}",
      journal = {\aap},
         year = 2005,
        month = sep,
       volume = {440},
       number = {3},
        pages = {981-994},
          doi = {10.1051/0004-6361:20053020},
archivePrefix = {arXiv},
       eprint = {astro-ph/0505229},
 primaryClass = {astro-ph},
       adsurl = {https://ui.adsabs.harvard.edu/abs/2005A&A...440..981T}
}

@ARTICLE{Moyano2023,
       author = {{Moyano}, F.~D. and {Eggenberger}, P. and {Salmon}, S.~J.~A.~J. and {Mombarg}, J.~S.~G. and {Ekstr{\"o}m}, S.},
        title = "{Angular momentum transport by magnetic fields in main-sequence stars with Gamma Doradus pulsators}",
      journal = {\aap},
         year = 2023,
        month = sep,
       volume = {677},
          eid = {A6},
        pages = {A6},
          doi = {10.1051/0004-6361/202346548},
archivePrefix = {arXiv},
       eprint = {2304.00674},
 primaryClass = {astro-ph.SR},
       adsurl = {https://ui.adsabs.harvard.edu/abs/2023A&A...677A...6M}
}

@ARTICLE{ChaboyerZahn1992,
       author = {{Chaboyer}, B. and {Zahn}, J. -P.},
        title = "{Effect of horizontal turbulent diffusion on transport by meridional circulation.}",
      journal = {\aap},
         year = 1992,
        month = jan,
       volume = {253},
        pages = {173-177},
       adsurl = {https://ui.adsabs.harvard.edu/abs/1992A&A...253..173C}
}

@ARTICLE{Mombarg2023-am,
       author = {{Mombarg}, J.~S.~G.},
        title = "{Calibrating angular momentum transport in intermediate-mass stars from gravity-mode asteroseismology}",
      journal = {\aap},
         year = 2023,
        month = sep,
       volume = {677},
          eid = {A63},
        pages = {A63},
          doi = {10.1051/0004-6361/202345956},
archivePrefix = {arXiv},
       eprint = {2306.17211},
 primaryClass = {astro-ph.SR},
       adsurl = {https://ui.adsabs.harvard.edu/abs/2023A&A...677A..63M}
}

@ARTICLE{Mombarg2024-gaia,
       author = {{Mombarg}, Joey S.~G. and {Aerts}, Conny and {Van Reeth}, Timothy and {Hey}, Daniel},
        title = "{Estimates of (convective core) masses, radii, and relative ages for {\ensuremath{\sim}}14 000 Gaia-discovered gravity-mode pulsators monitored by TESS}",
      journal = {\aap},
         year = 2024,
        month = nov,
       volume = {691},
          eid = {A131},
        pages = {A131},
          doi = {10.1051/0004-6361/202451651},
archivePrefix = {arXiv},
       eprint = {2410.05367},
 primaryClass = {astro-ph.SR},
       adsurl = {https://ui.adsabs.harvard.edu/abs/2024A&A...691A.131M}
}

@ARTICLE{Mombarg2024-distr,
       author = {{Mombarg}, J.~S.~G. and {Aerts}, C. and {Molenberghs}, G.},
        title = "{Probability distributions of initial rotation velocities and core-boundary mixing efficiencies of {\ensuremath{\gamma}} Doradus stars}",
      journal = {\aap},
         year = 2024,
        month = may,
       volume = {685},
          eid = {A21},
        pages = {A21},
          doi = {10.1051/0004-6361/202449213},
archivePrefix = {arXiv},
       eprint = {2402.05171},
 primaryClass = {astro-ph.SR},
       adsurl = {https://ui.adsabs.harvard.edu/abs/2024A&A...685A..21M}
}

@ARTICLE{Heger2005,
       author = {{Heger}, A. and {Woosley}, S.~E. and {Spruit}, H.~C.},
        title = "{Presupernova Evolution of Differentially Rotating Massive Stars Including Magnetic Fields}",
      journal = {\apj},
         year = 2005,
        month = jun,
       volume = {626},
       number = {1},
        pages = {350-363},
          doi = {10.1086/429868},
archivePrefix = {arXiv},
       eprint = {astro-ph/0409422},
 primaryClass = {astro-ph},
       adsurl = {https://ui.adsabs.harvard.edu/abs/2005ApJ...626..350H}
}

@ARTICLE{Aerts2025,
       author = {{Aerts}, Conny and {Van Reeth}, Timothy and {Mombarg}, Joey S.~G. and {Hey}, Daniel},
        title = "{Evolution of the near-core rotation frequency of 2497 intermediate-mass stars from their dominant gravito-inertial mode (Corrigendum)}",
      journal = {\aap},
         year = 2025,
        month = jun,
       volume = {698},
          eid = {C3},
        pages = {C3},
          doi = {10.1051/0004-6361/202555611e},
       adsurl = {https://ui.adsabs.harvard.edu/abs/2025A&A...698C...3A}
}

@ARTICLE{Li2024,
       author = {{Li}, Gang and {Aerts}, Conny and {Bedding}, Timothy R. and {Fritzewski}, Dario J. and {Murphy}, Simon J. and {Van Reeth}, Timothy and {Montet}, Benjamin T. and {Jian}, Mingjie and {Mombarg}, Joey S.~G. and {Gossage}, Seth and {Sreenivas}, Kalarickal R.},
        title = "{Asteroseismology of the young open cluster NGC 2516. I. Photometric and spectroscopic observations}",
      journal = {\aap},
         year = 2024,
        month = jun,
       volume = {686},
          eid = {A142},
        pages = {A142},
          doi = {10.1051/0004-6361/202348901},
archivePrefix = {arXiv},
       eprint = {2311.16991},
 primaryClass = {astro-ph.SR},
       adsurl = {https://ui.adsabs.harvard.edu/abs/2024A&A...686A.142L}
}

@ARTICLE{Aerts2025-AM,
       author = {{Aerts}, Conny},
        title = "{Distributions and evolution of the equatorial rotation velocities of 2937 BAF-type main-sequence stars from asteroseismology: A break in the specific angular momentum at M ≃ 2.5 M$_{{\ensuremath{\odot}}}$}",
      journal = {\aap},
         year = 2025,
        month = dec,
       volume = {704},
          eid = {A332},
        pages = {A332},
          doi = {10.1051/0004-6361/202556794},
archivePrefix = {arXiv},
       eprint = {2511.02909},
 primaryClass = {astro-ph.SR},
       adsurl = {https://ui.adsabs.harvard.edu/abs/2025A&A...704A.332A}
}

@ARTICLE{Fritzewski2024-gdor,
       author = {{Fritzewski}, D.~J. and {Aerts}, C. and {Mombarg}, J.~S.~G. and {Gossage}, S. and {Van Reeth}, T.},
        title = "{Age uncertainties of red giants due to cumulative rotational mixing of progenitors calibrated by asteroseismology}",
      journal = {\aap},
         year = 2024,
        month = apr,
       volume = {684},
          eid = {A112},
        pages = {A112},
          doi = {10.1051/0004-6361/202449300},
archivePrefix = {arXiv},
       eprint = {2402.05168},
 primaryClass = {astro-ph.SR},
       adsurl = {https://ui.adsabs.harvard.edu/abs/2024A&A...684A.112F}
}

@ARTICLE{Matt2015,
       author = {{Matt}, Sean P. and {Brun}, A. Sacha and {Baraffe}, Isabelle and {Bouvier}, J{\'e}r{\^o}me and {Chabrier}, Gilles},
        title = "{The Mass-dependence of Angular Momentum Evolution in Sun-like Stars}",
      journal = {\apjl},
         year = 2015,
        month = jan,
       volume = {799},
       number = {2},
          eid = {L23},
        pages = {L23},
          doi = {10.1088/2041-8205/799/2/L23},
archivePrefix = {arXiv},
       eprint = {1412.4786},
 primaryClass = {astro-ph.SR},
       adsurl = {https://ui.adsabs.harvard.edu/abs/2015ApJ...799L..23M}
}

@ARTICLE{Gallet2013,
       author = {{Gallet}, F. and {Bouvier}, J.},
        title = "{Improved angular momentum evolution model for solar-like stars}",
      journal = {\aap},
         year = 2013,
        month = aug,
       volume = {556},
          eid = {A36},
        pages = {A36},
          doi = {10.1051/0004-6361/201321302},
archivePrefix = {arXiv},
       eprint = {1306.2130},
 primaryClass = {astro-ph.SR},
       adsurl = {https://ui.adsabs.harvard.edu/abs/2013A&A...556A..36G}
}

@ARTICLE{Moyano2024,
       author = {{Moyano}, F.~D. and {Eggenberger}, P. and {Salmon}, S.~J.~A.~J.},
        title = "{Angular momentum transport near convective-core boundaries of Gamma Doradus stars}",
      journal = {\aap},
         year = 2024,
        month = jan,
       volume = {681},
          eid = {L16},
        pages = {L16},
          doi = {10.1051/0004-6361/202348704},
archivePrefix = {arXiv},
       eprint = {2401.05543},
 primaryClass = {astro-ph.SR},
       adsurl = {https://ui.adsabs.harvard.edu/abs/2024A&A...681L..16M}
}

@BOOK{Silverman1986,
       author = {{Silverman}, B. W.},
        title = "{Density Estimation for Statistics and Data Analysis}",
         year = 1986,
         publisher = "Monographs on Statistics and Applied Probability, Chapman and Hall, London",
         volume = {26}
}

@ARTICLE{Brun2005,
       author = {{Brun}, Allan Sacha and {Browning}, Matthew K. and {Toomre}, Juri},
        title = "{Simulations of Core Convection in Rotating A-Type Stars: Magnetic Dynamo Action}",
      journal = {\apj},
         year = 2005,
        month = aug,
       volume = {629},
       number = {1},
        pages = {461-481},
          doi = {10.1086/430430},
archivePrefix = {arXiv},
       eprint = {astro-ph/0610072},
 primaryClass = {astro-ph},
       adsurl = {https://ui.adsabs.harvard.edu/abs/2005ApJ...629..461B}
}

\end{document}